\documentstyle[preprint,aps,eqsecnum,tighten]{revtex}

\newtheorem{theorem}{Theorem}[section]
\newtheorem{definition}{Definition}[section]
\newtheorem{lemma}{Lemma}[section]

\def\be{\begin{equation}}
\def\ee{\end{equation}}
\def\ba{\begin{eqnarray}}
\def\ea{\end{eqnarray}}
\def\iff{\Leftrightarrow}
\def\<{\langle}
\def\>{\rangle}
\def\a{\alpha}
\def\b{\beta}
\def\c{\gamma}
\def\de{\delta}
\def\D{\Delta}
\def\d{{\mathrm d}}
\def\grad{\nabla}
\def\i{\mathrm{i}}
\def\C{{\mathbb C}}
\def\1{{\bf 1}}
\def\H{{\cal H}}
\def\punctH{{\cal H}^{\times}}
\def\P{{\cal P}}
\def\M{{\cal M}}
\def\w{\omega}
\def\W{\Omega}
\def\J{{\cal J}}

\def\implies{\Rightarrow}
\newcommand{\eqn}[1]{Eq.~(\ref{#1})}
\newcommand{\lie}[1]{{\pounds}_{#1}}
\newcommand{\hvf}[1]{{X_{#1}}}

\newtheorem{cor}{Corollary}
\def\Bbb{}

\def\Rl{{\mathchoice 
{\setbox0=\hbox{$\displaystyle\rm R$}\hbox{\hbox to0pt
{\kern0.4\wd0\vrule height0.9\ht0\hss}\box0}}
{\setbox0=\hbox{$\textstyle\rm R$}\hbox{\hbox to0pt
{\kern0.4\wd0\vrule height0.9\ht0\hss}\box0}}
{\setbox0=\hbox{$\scriptstyle\rm R$}\hbox{\hbox to0pt
{\kern0.4\wd0\vrule height0.9\ht0\hss}\box0}}
{\setbox0=\hbox{$\scriptscriptstyle\rm R$}\hbox{\hbox to0pt
{\kern0.4\wd0\vrule height0.9\ht0\hss}\box0}}}}
\def\Co{{\mathchoice
{\setbox0=\hbox{$\displaystyle\rm C$}\hbox{\hbox to0pt
{\kern0.4\wd0\vrule height0.9\ht0\hss}\box0}}
{\setbox0=\hbox{$\textstyle\rm C$}\hbox{\hbox to0pt
{\kern0.4\wd0\vrule height0.9\ht0\hss}\box0}}
{\setbox0=\hbox{$\scriptstyle\rm C$}\hbox{\hbox to0pt
{\kern0.4\wd0\vrule height0.9\ht0\hss}\box0}}
{\setbox0=\hbox{$\scriptscriptstyle\rm C$}\hbox{\hbox to0pt
{\kern0.4\wd0\vrule height0.9\ht0\hss}\box0}}}}

\def\Comp{{\mathchoice
{\setbox0=\hbox{$\displaystyle\rm C$}\hbox{\hbox to0pt
{\kern0.4\wd0\vrule height0.9\ht0\hss}\box0}}
{\setbox0=\hbox{$\textstyle\rm C$}\hbox{\hbox to0pt
{\kern0.4\wd0\vrule height0.9\ht0\hss}\box0}}
{\setbox0=\hbox{$\scriptstyle\rm C$}\hbox{\hbox to0pt
{\kern0.4\wd0\vrule height0.9\ht0\hss}\box0}}
{\setbox0=\hbox{$\scriptscriptstyle\rm C$}\hbox{\hbox to0pt
{\kern0.4\wd0\vrule height0.9\ht0\hss}\box0}}}}
\def\Co{{\mathchoice
{\setbox0=\hbox{$\displaystyle\rm C$}\hbox{\hbox to0pt
{\kern0.4\wd0\vrule height0.9\ht0\hss}\box0}}
{\setbox0=\hbox{$\textstyle\rm C$}\hbox{\hbox to0pt
{\kern0.4\wd0\vrule height0.9\ht0\hss}\box0}}
{\setbox0=\hbox{$\scriptstyle\rm C$}\hbox{\hbox to0pt
{\kern0.4\wd0\vrule height0.9\ht0\hss}\box0}}
{\setbox0=\hbox{$\scriptscriptstyle\rm C$}\hbox{\hbox to0pt
{\kern0.4\wd0\vrule height0.9\ht0\hss}\box0}}}}
\def\Rl{{\mathchoice
{\setbox0=\hbox{$\displaystyle\rm R$}\hbox{\hbox to0pt
{\kern0.4\wd0\vrule height0.9\ht0\hss}\box0}}
{\setbox0=\hbox{$\textstyle\rm R$}\hbox{\hbox to0pt
{\kern0.4\wd0\vrule height0.9\ht0\hss}\box0}}
{\setbox0=\hbox{$\scriptstyle\rm R$}\hbox{\hbox to0pt
{\kern0.4\wd0\vrule height0.9\ht0\hss}\box0}}
{\setbox0=\hbox{$\scriptscriptstyle\rm R$}\hbox{\hbox to0pt
{\kern0.4\wd0\vrule height0.9\ht0\hss}\box0}}}}

\def\R{\Rl} 
\def\C{\Comp}
\def\ch{{\cal C}_H}

\begin{document}

\pagenumbering{arabic}

\title{Geometrical Formulation of Quantum Mechanics}
\author{Abhay Ashtekar${}^{1,2}$ and Troy A. Schilling${}^{1,3}$}

\address{${}^1$ Center for Gravitational Physics and Geometry\\
Department of Physics, Penn State, 
University Park, PA 16802-6300, USA}
 
\address{${}^2$ Erwin Schr\"odinger International Institute for 
Mathematical Physics\\
Boltzmanngasse 9, A-1090 Vienna, Austria}

\address{${}^3$ Institute for Defense Analyses\\
1801 North Beauregard Street, Alexandria, VA 22311-1772}

\maketitle
\begin{abstract}

States of a quantum mechanical system are represented by rays in a
complex Hilbert space. The space of rays has, naturally, the structure
of a K\"ahler manifold. This leads to a geometrical formulation of the
postulates of quantum mechanics which, although equivalent to the
standard algebraic formulation, has a very different appearance.  In
particular, states are now represented by points of a symplectic
manifold (which happens to have, in addition, a compatible Riemannian
metric), observables are represented by certain real-valued functions
on this space and the Schr\"odinger evolution is captured by the
symplectic flow generated by a Hamiltonian function. There is thus a
remarkable similarity with the standard symplectic formulation of
classical mechanics. Features---such as uncertainties and state vector
reductions---which are specific to quantum mechanics can also be
formulated geometrically but now refer to the Riemannian metric---a
structure which is absent in classical mechanics. The geometrical
formulation sheds considerable light on a number of issues such as the
second quantization procedure, the role of coherent states in
semi-classical considerations and the WKB approximation.  More
importantly, it suggests generalizations of quantum mechanics. The
simplest among these are equivalent to the dynamical generalizations
that have appeared in the literature. The geometrical reformulation
provides a unified framework to discuss these and to correct a
misconception.  Finally, it also suggests directions in which more
radical generalizations may be found.  
\end{abstract}

\section{Introduction}\label{sec1}

Quantum mechanics is probably the most successful scientific theory
ever invented. It has an astonishing range of applications---from
quarks and leptons to neutron stars and white dwarfs---and the
accuracy with which its underlying ideas have been tested is equally
impressive. Yet, from its very inception, prominent physicists have
expressed deep reservations about its conceptual foundations and
leading figures continue to argue that it is incomplete in its
core. Time and again, attempts have been made to extend it in a
non-trivial fashion. Some of these proposals have been
phenomenological (see, e.g., \cite{grw,pearle1,ggr}), aimed at
providing a `mechanism' for the state reduction process. Some have
been more radical, e.g., invoking hidden variables (see, e.g.,
\cite{jb1}). Yet others involve non-linear generalizations of the
Schr\"odinger equation \cite{birula,pearle2,weinberg}. Deep discomfort
has been expressed at the tension between objective descriptions of
happenings provided by the space-time geometry of special relativity
and the quantum measurement theory \cite{pearle3,jb2}. Further
conceptual issues arise when one brings general relativity in to
picture, issues that go under the heading of `problem of time' in
quantum gravity \cite{as,kk,ci1}. Thus, while there is universal
agreement that quantum mechanics is an astonishingly powerful working
tool, in the `foundation of physics circles' there has also been a
strong sentiment that sooner or later one would be forced to
generalize it in a profound fashion \cite{rp1,gh,ci2}.

It is often the case that while an existing theory admits a number of
equivalent descriptions, one of them suggests generalizations more
readily than others. Furthermore, typically, this description is not
the most familiar one, i.e., not the one that seems simplest from the
limited perspective of the existing theory. An example is provided by
Cartan's formulation of Newtonian gravity. While it played no role in
the invention of the theory (it came some two and a half centuries
later!) at a conceptual level, Cartan's framework provides a deeper
understanding of Newtonian gravity and its relation to general
relativity. A much more striking example is Minkowski's geometric
reformulation of special relativity.  His emphasis on hyperbolic
geometry seemed abstract and abstruse at first; at the time, Einstein
himself is said to have remarked that it made the subject
incomprehensible to physicists. Yet, it proved to be an essential
stepping stone to general relativity.

The purpose of this article is to present, in this spirit, a
reformulation of the mathematical framework underlying standard
quantum mechanics (and quantum field theory). The strength of the
framework is that it is extremely natural from a geometric perspective
and succinctly illuminates the essential difference between classical
and quantum mechanics. It has already clarified certain issues related
to the second quantization procedure and semi-classical approximations
\cite{thesis}. It also serves to unify in a coherent fashion a number
of proposed generalizations of quantum mechanics; in particular, we
will see that generalizations that were believed to be distinct (and
even incompatible) are in fact closely related. More importantly, this
reformulation may well lead to viable generalizations of quantum
mechanics which are more profound than the ones considered so
far. Finally, our experience from seminars and discussions has shown
that ideas underlying this reformulation lie close to the heart of
geometrically oriented physicists. It is therefore surprising that the
framework is not widely known among relativists. We are particularly
happy to be able to rectify this situation in this volume honoring
Engelbert and hope that the role played by the K\"ahler geometry, in
particular, will delight him.

Let us begin by comparing the standard frameworks underlying classical
and quantum mechanics. The classical description is {\it geometrical}:
States are represented by points of a symplectic manifold $\Gamma$,
the phase space.  The space of observables consists of the (smooth)
real-valued functions on this manifold. The (ideal) measurement of an
observable $f$ in a state $p \in \Gamma$ yields simply the value
$f(p)$ at the point $p$; the state is left undisturbed. These outcomes
occur with complete certainty. The space of observables is naturally
endowed with the structure of a commutative, associative algebra, the
product being given simply by pointwise multiplication. Thanks to the
symplectic structure, it also inherits a Lie-bracket---the Poisson
bracket. Finally to each observable $f$ is associated a vector field
$X_f$ called the Hamiltonian vector field of $f$. Thus, each
observable generates a flow on $\Gamma$. Dynamics is determined by a
preferred observable, the Hamiltonian $H$; the flow generated by $X_H$
describes the time evolution of the system.

The arena for quantum mechanics, on the other hand, is a Hilbert space
$\H$.  States of the system now correspond to rays in $\H$, and the
observables are represented by self-adjoint linear operators on
$\H$. As in the classical description, the space of observables is a
real vector space equipped with with two algebraic structures. First,
we have the the Jordan product---i.e., the anti-commutator---which is
commutative but now fails to be associative. Second, we have ($1/2i$
times) the commutator bracket which endows the space of observables
with the structure of a Lie algebra.  Measurement theory, on the other
hand, is strikingly different. In the textbook description based on
the Copenhagen interpretation, the (ideal) measurement of an
observable $\hat{A}$ in a state $\Psi \in \H$ yields {\it an
eigenvalue} of $\hat{A}$ and, immediately after the measurement, the
state is thrown into the corresponding eigenstate.
specific outcome can only be predicted probabilistically. As in the
classical theory, each observable $\hat{A}$ gives rise to a flow on
the state space. But now, the flow is generated by the 1-parameter
group $\exp i\hat{A}t$ and respects the linearity of $\H$.  Dynamics
is again dictated by a preferred observable, the Hamiltonian operator
$\hat{H}$.

Clearly, the two descriptions have several points in common.  However,
there is also a striking difference: While the classical framework is
{\it geometric and non-linear}, the quantum description is
intrinsically {\it algebraic and linear}. Indeed, the emphasis on the
underlying linearity is so strong that none of the standard textbook
postulates of quantum mechanics can be stated without reference to the
linear structure of $\H$.

{}From a general perspective, this difference seems quite surprising.
For linear structures in physics generally arise as approximations to
the more accurate non-linear ones. Thus, for example, we often
encounter non-linear equations which correctly capture a physical
situation. But, typically, they are technically difficult to work with
and we probe properties of their solutions through linearization.  In
the present context, on the other hand, it is the deeper, more correct
theory that is linear and the non-linear, geometric, classical
framework is to arise as a suitable limiting case.

However, deeper reflection shows that quantum mechanics is in fact not
as linear as it is advertised to be. For, the space of physical states
is {\it not} the Hilbert space $\H$ but the space of rays in it, i.e.,
the {\it projective} Hilbert space $\P$. And $\P$ is a genuine,
non-linear manifold. Furthermore, it turns out that the Hermitian
inner-product of $\H$ naturally endows $\P$ with the structure of a
K\"ahler manifold. Thus, in particular, like the classical state space
$\Gamma$, the correct space of quantum states, $\P$, is a symplectic
manifold! We will therefore refer to $\P$ as the {\it quantum phase
space}. Given any self-adjoint operator $\hat{H}$, we can take its
expectation value to obtain a real function on $\H$. It is easy to
verify that this function admits an unambiguous projection $h$ to the
projective Hilbert space $\P$. Recall, now, that every phase space
function gives rise to a flow through its Hamiltonian vector field.
What then is the interpretation of the flow $X_h$? It turns out
\cite{kibble} to be exactly the (projection to $\P$ of the) flow
defined by the Schr\"odinger equation (on $\H$) of the quantum theory.
Thus, Schr\"odinger evolution is precisely the Hamiltonian flow on the
quantum phase space!

As we will see, the interplay between the classical and quantum ideas
stretches much further. The overall picture can be summarized as
follows. Classical phase spaces $\Gamma$ are, in general, equipped
only with a symplectic structure.  Quantum phase spaces, $\P$, on the
other hand, come with an additional structure, the Riemannian metric
provided by the K\"ahler structure.  Roughly speaking, features of
quantum mechanics which have direct classical analogues refer only to
the symplectic structure.  On the other hand, features---such as
quantum uncertainties and state vector reduction in a measurement
process---refer also the Riemannian metric. This neat division lies at
the heart of the structural similarities and differences between the
(mathematical frameworks underlying the) two theories.

Section \ref{sec2} summarizes this geometrical reformulation of
standard quantum mechanics. We begin in \ref{sec2.A} by showing that
the quantum Hilbert space can be regarded as a (linear) K\"ahler space
and discuss the roles played by the symplectic structure and the
K\"ahler metric. In \ref{sec2.B}, we show that one can naturally
arrive at the quantum state space ${\cal P}$ by using the
Bergmann-Dirac theory of constrained systems. (This method of
constructing the quantum phase space will turn out to be especially
convenient in section \ref{sec3} while analyzing the relation between
various generalizations of quantum mechanics.) Section \ref{sec2.C}
provides a self-contained treatment of the various issues related to
observables---associated algebraic structures, quantum uncertainty
relations and measurement theory---in an intrinsically geometric
fashion. These results are collected in section \ref{sec2.D} to obtain
a geometric formulation of the postulates of quantum mechanics, {\it a
formulation that makes no reference to the Hilbert space $\H$ or the
associated linear structures}. In practical applications, except while
dealing with simple cases such as spin systems, the underlying quantum
phase space $\P$ is {\it infinite}-dimensional (since it comes from an
infinite-dimensional Hilbert space $\H$). {\it Our mathematical
discussion encompasses this case}. Also, in the discussion of
measurement theory, we allow for the possibility that observables may
have continuous spectra.

In section \ref{sec3}, we consider possible generalizations of quantum
mechanics. These generalizations can occur in two distinct
ways. First, we can retain the original kinematic structure but allow
more general dynamics, e.g., by replacing the Schr\"odinger equation
by a suitable non-linear one (see, e.g., \cite{birula}).  In section
\ref{sec3.A} we show that the geometrical reformulation of section
\ref{sec2} naturally suggests a class of such extensions which
encompasses those proposed by Birula and Mycielski \cite{birula} and
by Weinberg \cite{weinberg}. In section \ref{sec3.B} we consider the
possibility of more radical extensions in which the kinematical set up
itself is changed.  Although (to our knowledge) there do not exist
interesting proposals of this type, such generalizations would be much
more interesting.  In particular, it is sometimes argued that the
linear structure underlying quantum mechanics would have to be
sacrificed in a subtle but essential way to obtain a satisfactory
quantum theory of gravity and/or to cope satisfactorily with the
`measurement problem' \cite{rp1}. To implement such ideas, the
underlying kinematic structure will have to be altered.  A first step
in this direction is to obtain a useful {\it characterization} of the
kinematical framework of standard quantum mechanics. Section
\ref{sec3.B} provides a `reconstruction theorem' which singles out
quantum mechanics from its plausible generalizations. The theorem
provides powerful guidelines: it spells out directions along which one
can proceed to obtain a genuine extension.

Section \ref{sec4} is devoted to semi-classical issues. Consider a
simple mechanical system, such as a particle in $\R^3$. In this case,
the classical phase space $\Gamma$ is six-dimensional while the
quantum phase space $\P$ is infinite-dimensional. Is there a relation
between the two? In section \ref{sec4.A} we show that the answer is in
the affirmative: $\P$ is a bundle over $\Gamma$. Furthermore, the
bundle is trivial. Thus, through each quantum state $p\in \P$, there
is a cross-section, i.e. a copy of $\Gamma$. It turns out that the
quantum states that lie on any one cross-section are precisely the
{\it generalized coherent states} \cite{perelomov,gilmore,klauder}. In
the remainder of section \ref{sec4}, we use this interplay between
$\P$ and $\Gamma$ to discuss the relation between classical and
quantum dynamics.  Section \ref{sec4.B} is devoted to the
correspondence in terms of Ehrenfest's theorem while \ref{sec4.C}
discusses the problem along the lines of the WKB
approximation. Somewhat interestingly, it turns out that WKB dynamics
is an example of generalized dynamics of the Weinberg type
\cite{weinberg}.

Our conventions are as follows. If the manifold under consideration is
infinite-dimensional, we will assume that it is a Hilbert
manifold. (Projective Hilbert spaces are naturally endowed with this
structure; see, e.g., \cite{thesis}.) Riemannian metrics and
symplectic structures on these manifolds will be assumed to be
everywhere defined, smooth, strongly non-degenerate fields. (Thus,
they define {\it isomorphisms} between the tangent and cotangent
spaces at each point). In detailed calculations we will often use the
abstract index notation of Penrose's \cite{rp2,indices}.  Note that,
in spite of the appearance of indices, this notation is well-defined
also on infinite-dimensional manifolds. (For example, if $V^a$ denotes
a contravariant vector field; the subscript $a$ does not refer to its
components but is only a label telling us that $V$ is a specific type
of tensor field, namely a contravariant vector field. Similarly,
$V^a\omega_a$ is the function obtained by the action of the 1-form
$\omega$ on the contravariant vector field $V$.)  Finally, due to
space limitation, we have not included detailed proofs of several
technical assertions; they can be found in \cite{thesis}. Our aim here
is only to provide a thorough overview of the subject.

This work was intended to be an extension of a paper by Kibble
{}\cite{kibble} which pointed out that the Schr\"odinger evolution can
be regarded as an Hamiltonian flow on $\H$. However, after completing
this work, we learned that many of the results contained in sections
\ref{sec2} and \ref{sec3.B} were obtained independently by others
(although the viewpoints and technical proofs are often distinct.)  In
1985, Heslot \cite{heslot} observed that quantum mechanics admits a
symplectic formulation in which the phase space is the projective
Hilbert space. That discussion was, however, restricted to the
finite-dimensional case and did not include a discussion of the role
of the metric, probabilistic interpretation and quantum uncertainties.
Anandan and Aharonov \cite{anandan} rediscovered some of these results
and also discussed some of the probabilistic aspects. This work was
also restricted to finite-dimensional systems and focussed on the
issue of evolution. Similar observations were made by Gibbons
\cite{gibbons} who also discussed density matrices (which are not
considered here) and raised the issue of characterization of quantum
mechanics (which is resolved in section \ref{sec4.B}). An essentially
complete treatment of the finite-dimensional case was given by
Hughston \cite{hughston}. (This work was done in parallel to
ours. However, it also contains some proposals for mechanisms for
state reduction \cite{hughston2} which are not discussed here.)  The
only references (to our knowledge) which treat the
infinite-dimensional case are \cite{italy1,italy2}, which also discuss
the issue of characterization of standard quantum mechanics. Finally,
since the geometric structures that arise here are so natural, it is
quite possible that they were independently discovered by other
authors that we are not aware of.

\section{Geometric formulation of quantum mechanics}\label{sec2}

The goal of this section is to show that quantum mechanics can be
formulated in an intrinsically geometric fashion, without any
reference to a Hilbert space or the associated linear structure.  We
will assume that the reader is familiar with basic symplectic
geometry.

\subsection{The Hilbert space as a K\"ahler space}
\label{sec2.A}

Let us begin with the standard Hilbert space formulation of quantum
mechanics.  In this sub-section we will view the Hilbert space $\H$ as
a K\"ahler space and examine the role played by the associated
symplectic structure and the Riemannian metric.  This discussion will
serve as a stepping stone to the analysis of the quantum phase space
$\P$ in section \ref{sec2.B}.

The similarities between classical and quantum mechanics can be put in
a much more suggestive form with an alternative, but equivalent,
description of the Hilbert space.  We view $\H$ as a {\em real} vector
space equipped with a {\em complex structure} $J$.  The complex
structure is a preferred linear operator which represents
multiplication by $\i$; hence $J^2 = -{I}$.  Initially, this change of
notation seems rather trivial; the element which is typically written
$(a + \i b)\Psi$ is now denoted $a\Psi + b J\Psi$ and (external)
multiplication of vectors by complex numbers is not permitted.
However, this slight change of viewpoint will come with a reward---a
symplectic formulation of quantum mechanics.

Since $\H$ is now viewed as a real vector space, the Hermitian
inner-product is slightly unnatural.  We therefore decompose it into
real and imaginary parts,
\be \< \Phi, \Psi\>
=: \frac{1}{2\hbar}G(\Phi, \Psi) + \frac{\i}{2\hbar}\W(\Phi, \Psi).
\ee 
(The reason for the factors of $1/2\hbar$ will become clear shortly.)
Properties of the Hermitian inner-product imply that $G$ is a positive
definite, real inner-product and that $\W$ is a symplectic form, both
of which are strongly non-degenerate.  Moreover, since $\<\Phi, J
\Psi\> = \i \<\Phi, \Psi\>$, one immediately observes that the metric,
symplectic structure and complex structure are related as
\be
\label{kahler_reln} G(\Phi, \Psi) = \W(\Phi, J \Psi).  
\ee 
That is, the triple $(J, G, \W)$ equips $\H$ with the structure of a
{\em K\"ahler space}.  Therefore, every Hilbert space may be naturally
viewed as a K\"ahler space.

Next, by use of the canonical identification of the tangent space (at
any point of $\H$) with $\H$ itself, $\W$ is naturally extended to a
strongly non-degenerate, closed, differential 2-form $\H$, which we
will denote also by $\W$.  Any Hilbert space is therefore naturally
viewed as the simplest sort of symplectic manifold, i.e., a {\em phase
space}.  The inverse of $\W$ may be used to define Poisson brackets
and Hamiltonian vector fields.  As we are about to see, these notions
are just as relevant in quantum mechanics as in classical mechanics.

\subsubsection{The symplectic form}

In classical mechanics, observables are real-valued functions, and to
each such function is associated a corresponding Hamiltonian vector
field.  In quantum mechanics, on the other hand, the observables
themselves may be viewed as vector fields, since linear operators
associate a vector to each element of the Hilbert space.  However, the
Schr\"odinger equation, which in our language is written as $ \dot\Psi
= - \frac{1}{\hbar} J \hat{H} \Psi, $ motivates us to associate to
each quantum observable $\hat{F}$ the vector field 
\be
\label{schrodinger_vf}
      Y_{\hat{F}} (\Psi) := -\frac{1}{\hbar} J \hat{F} \Psi.
\ee
This {\em Schr\"odinger vector field} is defined so that the
time-evolution of the system corresponds to the flow along the
Schr\"odinger vector field associated to the Hamiltonian operator.
(Note that, if the Hamiltonian is unbounded, it is only densely
defined and so is the vector field. The (unitary) flow, however, is
defined on all of $\H$. See, e.g., \cite{chernoff_marsden}.)

Natural questions immediately arise.  Let $\hat{F}$ be any bounded,
self-adjoint operator on $\H$.  Is the corresponding vector field,
$Y_{\hat{F}}$, Hamiltonian on the symplectic space $(\H, \Omega)$?  If
so, what is the real-valued function which generates this vector
field?  What is the physical meaning of the Poisson bracket?  In
particular, how is it related to the commutator Lie algebra?

The answers to these questions are remarkably simple.  As we know from
standard quantum mechanics, $\hat{F}$ generates a one-parameter family
of unitary mappings on $\H$.  By definition, $Y_{\hat{F}}$ is the
generator of this one-parameter family and therefore preserves both
the metric $G$ and symplectic form $\W$.  It is therefore locally
Hamiltonian, and, since $\H$ is a linear space, also globally
Hamiltonian!  In fact, the function which generates this Hamiltonian
vector field is of physical interest; it is simply the expectation
value of $\hat{F}$.

Let us see this explicitly.  Denote by $F : \H \longrightarrow \R$ the
expectation value function,
\be
	F(\Psi) := \< \Psi, \hat{F}\Psi \> 
		 = \frac{1}{2\hbar}G(\Psi, \hat{F}\Psi).
\ee
We will continue to use this notation; expectation value functions
will be denoted by simply ``un-hatting'' the corresponding operators.
Now, if $\eta$ is any tangent vector at $\Psi$, then%
\footnote{With our conventions for symplectic geometry, the
Hamiltonian vector field $\hvf{f}$ generated by the function $f$
satisfies the equation $i_\hvf{f}\W = \d f$, and the Poisson bracket
is defined by $\{f, g \} = \W(\hvf{f}, \hvf{g})$.}
\ba
\label{SVF_is_HVF}
(\d F)(\eta) &= \frac{\d}{\d t} \< \Psi + t\eta \hat{F}
(\Psi + t \eta) \> \big|_{t=0}
= \< \Psi, \hat{F}\eta \> + \< \eta, \hat{F}\Psi \>\nonumber \\
&= \frac{1}{\hbar} G(\hat{F}\Psi, \eta)
= \W(Y_{\hat{F}}, \eta) = \big(i_{Y_{\hat{F}}} \W\big)(\eta),
\ea
where we have used the self-adjointness of $\hat{F}$,
\eqn{kahler_reln} and the definition of $Y_{\hat{F}}$.  Therefore, the
Hamiltonian vector field $\hvf{F}$ generated by the expectation value
function $F$ coincides with the Schr\"odinger vector field
$Y_{\hat{F}}$ associated to $\hat{F}$.  As a particular consequence,
the time evolution of any quantum mechanical system may be written in
terms of Hamilton's equation of classical mechanics; the Hamiltonian
{\em function} is simply the expectation value of the Hamiltonian {\em
operator}.  {\em Schr\"odinger's equation is Hamilton's equation in
disguise!}

Next, let $\hat{F}$ and $\hat{K}$ be two quantum observables, and
denote by $F$ and $K$ the respective expectation value functions.  It
is natural to ask whether the Poisson bracket of $F$ and $K$ is
related in a simple manner to an algebraic operation involving the
original operators. Performing a calculation as simple as
\eqn{SVF_is_HVF}, one finds that
\be \label{COMMUTATOR_is_PB}
\{ F, K \}_\W = \W(\hvf{F}, \hvf{K})
= \bigg< \frac{1}{\i\hbar} [\hat{F}, \hat{K}] \bigg>.
\ee
Notice that the quantity inside the brackets on the right side of
\eqn{COMMUTATOR_is_PB} is precisely the quantum Lie bracket of
$\hat{F}$ and $\hat{K}$.  The algebraic operation on the expectation
value functions, which is induced by the commutator bracket is {\em
exactly} a Poisson bracket!  Note that this is {\em not} Dirac's
correspondence principle; the Poisson bracket here is the quantum one,
determined by the imaginary part of the Hermitian inner-product.

The basic features of the classical formalism appear also in quantum
mechanics.  The Hilbert space, as a real vector space, is equipped
with a symplectic form.  To each quantum observable is associated a
real-valued function on $\H$, and the time-evolution is determined by
the Hamiltonian vector field associated to a preferred function.
Moreover, the Lie bracket of two quantum observables corresponds
precisely to the Poisson bracket of the corresponding functions.

\subsubsection{Uncertainty and the real inner-product}

Let us now examine the role played by the metric $G$. Clearly, $G$
enables us to define a real inner-product, $G(X_F, X_K)$ between any
two Hamiltonian vector fields $X_F$ and $X_K$. One may expect that
this inner-product is related to the Jordan product in much the same
way that the symplectic form corresponds to the commutator Lie
bracket. It is easy to verify that this expectation is
correct. Operating just as in \eqn{COMMUTATOR_is_PB}, we obtain 
\be
\label{symmetric_bracket}
\{ F, K \}_+ := \frac{\hbar}{2} G(\hvf{F}, \hvf{K})
= \bigg< \frac{1}{2} [\hat{F}, \hat{K}]_+ \bigg>.
\ee
The operation defined by the first equation above will be called the
{\em Riemann bracket} of $F$ and $K$.  Up to the factor of
$\hbar / 2$, the Riemann bracket of $F$ and $K$ is simply given
by the (real) inner-product of their Hamiltonian vector fields, and
corresponds precisely to the Jordan product of the respective
operators.

Since the classical phase space is, in general, not equipped with a
Riemannian metric, the Riemann product does {\em not} have an analogue
in the classical formalism; it does, however, admit a physical
interpretation.  In order to see this, note that the uncertainty of
the observable $\hat{F}$ at a state with unit norm is given by
\be \label{uncertainty}
(\D \hat{F})^2 = \< \hat{F}^2 \> - \< \hat{F} \>^2
= \{ F, F \}_+ - F^2.
\ee
Thus, the uncertainty of an operator $\hat{F}$, when written in terms
of the expectation value function $F$, involves the Riemann bracket.
Moreover, this expression for the uncertainty is quite simple.

In fact Heisenberg's famous uncertainty relation also assumes a nice
form when expressed in terms of the expectation value functions. It is
very well-known that the familiar uncertainty relation between two
quantum observables may be written in a slightly stronger
form (see, e.g.,\cite{shankar}): 
\be \label{std_unc_reln}
(\D \hat{F})^2 (\D \hat{K})^2 \ge
\bigg< \frac{1}{2\i} [ \hat{F}, \hat{K} ] \bigg>^2
+ \bigg< \frac{1}{2} [ \hat{F}_{\perp}, 
\hat{K}_{\perp} ]_+ \bigg>^2,
\ee
where $\hat{F}_{\perp}$ is the {\em non-linear} operator defined
by
\[
 \hat{F}_{\perp}(\Psi) := \hat{F}(\Psi) - F(\Psi)
\]
so that $\hat{F}_{\perp}(\Psi)$ is orthogonal to $\Psi$,
if $\| \Psi \| = 1$.

Using the above results, we may immediately rewrite \eqn{std_unc_reln}
in the form 
\be \label{new_unc_reln}
(\D \hat{F})^2 (\D \hat{K})^2 \ge
\bigg( \frac{\hbar}{2} \{ F, K \}_\W \bigg)^2
+ \big( \{ F, K \}_+ - FK \big)^2.
\ee
Incidentally, the last expression in \eqn{new_unc_reln} may be
interpreted as the ``quantum covariance'' of $\hat{F}$ and $\hat{K}$;
see Ref. \onlinecite{thesis} for an explanation.

\subsection{The quantum phase space} \label{sec2.B}

We have seen that to each quantum observable is associated a smooth
real-valued (expectation value) function on the Hilbert space.
Further, the familiar operations involving quantum operators
correspond to simple ``classical looking'' operations on the
corresponding functions.  These observations suggest a formulation of
standard quantum mechanics in the language of classical mechanics.
However, there are two difficulties.  First, although the Hilbert
space is a symplectic space, because two state vectors related by
multiplication by any complex number define the same state, it is {\it
not} the space of physical states, i.e., the quantum analog of the
classical phase space.  Second, the description of the measurement
process in a manner intrinsic to the K\"ahler structure on $\H$ turns
out not to be natural.

The true state space of the quantum system is the space of {\em rays}
in the Hilbert space, i.e., the {\em projective} Hilbert space, which
we shall denote $\P$.  It should not be surprising that $\P$ is a
K\"ahler manifold, and hence, in particular, a symplectic manifold.
After all, for the special case in which $\H$ is $\C^{n+1}$, $\P$ is
the complex projective space ${\Bbb CP}^n$---the archetypical K\"ahler
manifold.  In this section, we present a particularly useful
description of the projective Hilbert space which is valid for the
{\it infinite-dimensional case} and which illuminates the role of its
symplectic structure.  These developments will enable us to handle the
above complications and allow an elegant geometric formulation of
quantum mechanics.

\subsubsection{Gauge reduction}

The standard strategy to handle the ambiguity of the state vector is
to consider only those elements of the Hilbert space which are
normalized to unity. We will adopt this approach by insisting that the
only physically relevant portion of the Hilbert space is that on which
the {\em constraint function},
\be \label{constraint}
C(\Psi) := \< \Psi, \Psi \> -1
= \frac{1}{2\hbar} G( \Psi, \Psi ) - 1,
\ee
vanishes.  The attitude adopted here is one in which the the {\em
constraint surface}---the unit sphere, with respect to the Hermitian
inner-product---is the only portion of the Hilbert space which is
accessible to the system.  The rest of the Hilbert space is often
quite convenient, but is viewed as an artificial element of the
formalism.

Let us consider the above restriction from the point of view of the
Bergmann-Dirac theory of constrained systems (see, e.g.,
\cite{bd}). In other words, we will pretend, for a moment, that we are
dealing with a {\em classical} theory with the constraint $C=0$.  First
notice that since the time-evolution preserves the constraint surface,
no further (secondary) constraints arise.  Since we have only
a single constraint, it is trivially of {\em first-class} in Dirac's
terminology; i.e., the constraint generates a motion which preserves
the constraint surface ($\lie{\hvf{C}} C = \{ C, C\}_\W = 0$).

Recall that to every first-class constraint on a Hamiltonian system is
associated a gauge degree of freedom; the associated gauge
transformations are defined by the flow along the Hamiltonian vector
field generated by the constraint function.  In our case, the gauge
directions are simply given by
\be X^a_{C} = \W^{ab} D_b C =
\frac{1}{\hbar}\W^{ab} \Psi_b
		  = - \frac{1}{\hbar} J^a{}_b \Psi^b,
\ee
where $D_a$ denotes the Levi-Civit\`a derivative operator.  For later
convenience, let us define
\be
	\J^a := \hbar \hvf{C}^a\big|_S = - J^a{}_b \Psi^b\big|_S.
\ee
Notice that $\J^a$ is the generator of phase rotations on $S$.
Therefore, the gauge transformations generated by the constraint are
exactly what they ought to be; they represent the arbitrariness in our
choice of phase!

Thus, we now see the relevance of the description in terms of
constrained Hamiltonian systems.  By taking the quotient of the
constraint surface of any constrained system by the action of the
gauge transformations, one obtains the true phase space of the
system---often called the {\em reduced phase space}.  The projective
Hilbert space may therefore be interpreted as the ``reduced phase
space'' of our constrained Hamiltonian system.  In order to emphasize
both its physical role and geometric structure, we will refer to the
projective Hilbert space $\P$ as the {\em quantum phase space}. One
can explicitly show that if $\H$ is infinite-dimensional, $\P$ is an
infinite-dimensional Hilbert manifold \cite{thesis}.

As the terminology suggests, any reduced phase space is equipped with
a natural symplectic structure.  This fact may be seen as follows.
Denote by $i:S\rightarrow\H$ and $\pi:S\rightarrow\P$ the inclusion
mapping and projection to the quantum phase space, respectively.  By
restricting the symplectic structure $\W$ to the constraint surface,
one obtains a closed 2-form $i^*\W$ on $S$.  This 2-form is
degenerate, but only in the gauge direction.  Fortunately, since gauge
transformations are defined by the Hamiltonian vector field generated
by the constraint function, $i^*\W$ is constant along its directions
of degeneracy.  As a result, there exists a symplectic form $\w$ on
$\P$ whose pull-back via $\pi$ agrees precisely with $i^*\W$. This is
a standard construction in the theory of systems with first class
constraints; the only novelty here lies in its application to ordinary
quantum mechanics.  Finally, we note that this result applies to the
typical case of interest, in which the original Hilbert space is
infinite-dimensional \cite{thesis}. The symplectic structure $\omega$
is then a smooth, strongly non-degenerate field (i.e., defines an {\it
isomorphism} from the tangent space to the cotangent space at each
point.)

Before discussing the geometry the quantum phase space, we should
point out that the viewpoint adopted in this section in fact
generalizes quantum mechanics, but in a very trivial way.  In section
\ref{sec2.A} we observed that each quantum observable defines a
real-valued function on the entire Hilbert space, and that the quantum
evolution is given by the Hamiltonian flow defined by a preferred
function.  In viewing a quantum system as a constrained Hamiltonian
system, we must concede that it is only the constraint surface $S$
that is of physical relevance.  In particular, the restriction $F|_S$
of the expectation value function $F$ contains all gauge-invariant
information about the observable; one may extend $F$ off the
constraint surface in any desirable manner without affecting the
corresponding flow on the projective space.  The particular extensions
defined by expectation values of (bounded) self-adjoint operators may
be viewed as mere convention.  We will make use of this point in
section \ref{sec3.A}.

\subsubsection{Symplectic geometry}

Our method of arriving at the quantum phase space $\P$ by the reduced
phase space construction of constrained systems immediately suggests
further definitions and constructions.

Recall that to each bounded, self-adjoint operator $\hat{F}$ on $\H$,
we have associated the function $F(\Psi) := \< \Psi, \hat{F} \Psi \>$
on the Hilbert space.  In fact, we may go a short step further.
First, let us restrict the expectation value function $F$ to the
constraint surface, thereby obtaining the function $i^*F : S
\rightarrow \R$.  $i^*F$ is clearly gauge-invariant (i.e. independent
of phase), and therefore defines the function $f: \P \rightarrow \R$
for which $\pi^*f = i^*F$.  Therefore, to each quantum observable is
associated a smooth, real-valued function on the quantum phase space.
The functions obtained in this manner will represent the observables
in the geometric formulation of quantum mechanics. Let us therefore
make
\begin{definition} \label{defn_observable_fn}
Let $f: \P \rightarrow \R$ be a smooth function on $\P$.  If there
exists a bounded, self-adjoint operator $\hat{F}$ on $\H$ for which
$\pi^*f = \< \hat{F} \>\big|_S$, then $f$ is said to be an {\em
observable function}.
\end{definition}
\noindent
Note that we consider the set of quantum observables to consist of the
{\em bounded} self-adjoint operators on $\H$. At first sight this
appears to be a severe restriction. However, further reflection shows
that it is not. In any actual experiment, one deals only with a finite
range of relevant parameters and hence in practice one only measures
observables of the type considered here.  Thus, there is by definition
a one-to-one correspondence between quantum observables and the
observable functions on $\P$.  As we will see below, the set of
observable functions is a very small subset of the entire function
space.

A natural question arises: What is the relationship between the
Hamiltonian vector fields $\hvf{F}$ (on $\H$) and $\hvf{f}$ (on $\P$)?
Given any point $\Psi \in S$, we may push-forward the vector
$\hvf{F}\big|_\Psi$ to obtain a tangent vector at $\pi(\Psi)$.  Since
$F$ is gauge-invariant, it commutes with $C$; therefore
\[	\lie{\hvf{C}} \hvf{F} = \hvf{ \{ F, C\}_\W} \equiv 0.	\]
As a consequence, $\hvf{F}$ is ``constant along the integral curves of
$\J$''.  Thus, by pushing-forward $\hvf{F}$ at each point of $S$, one
obtains a well-defined (smooth) vector field on all of $\P$.  As is
known to those familiar with the analysis of constrained systems, this
vector field is also Hamiltonian; in fact, it agrees precisely with
$\hvf{f}$.  The flow on $\P$, which is induced by the Schr\"odinger
vector field of $\hat{F}$ corresponds exactly to the Hamiltonian flow
determined by the observable function $f$.

Next, consider the Poisson bracket $\{ , \}_\w$ defined by the reduced
symplectic structure $\w$.  Let $F, K :\H \rightarrow \R$ be
expectation value functions of two quantum observables and denote by
$f,k:\P \rightarrow \R$ the corresponding observable functions on
$\P$.  As a consequence of the above result,
\be \label{pb_on_P}
\pi^*\{ f, k \}_\w = \pi^*( \w( \hvf{f}, \hvf{k} ) )
= \w( \pi_*\hvf{F}, \pi_*\hvf{K} )
= \W( \hvf{F}, \hvf{K} )\big|_S
= \{ F, K \}_\W\big|_S.
\ee
Therefore, the Poisson bracket defined by $\w$ {\em exactly} reflects
the commutator bracket on the space of quantum observables.

In summary, to each quantum observable $\hat{F}$ is associated a
real-valued function $f:\P \rightarrow \R$ on the quantum phase space.
The Schr\"odinger vector field determined by $\hat{F}$ determines a
flow on $\P$; this flow is generated by the Hamiltonian vector field
associated to the observable function $f$.  Further, the mapping
$\hat{F} \mapsto f$ is one-to-one and respects the Lie algebraic
structures provided by the commutator and Poisson bracket on $\P$,
respectively.

\subsection{Riemannian geometry and measurement theory}
\label{sec2.C}

Any quantum mechanical system may be described as an
infinite-dimensional Hamiltonian system.  However, the structure of
the quantum phase space is much richer than that of classical
mechanics.  $\P$ is also equipped with a natural Riemannian metric.
As we will see, the probabilistic features of quantum mechanics are
conveniently described by the Riemannian structure.

The quantum metric may be described in much the same way as the
symplectic structure.  The restriction $i^*G$ of $G$ to the unit
sphere is a strongly non-degenerate Riemannian metric on $S$.  Recall
that the gauge generator $\J$ is (up to the constant factor of
$\hbar$) the Schr\"odinger vector field associated to the identity
operator.  Since any Schr\"odinger vector field preserves the
Hermitian inner-product, it preserves both the symplectic structure
$\W$ and the metric $G$.  As a consequence, $\J$ is a Killing vector
field on $S$;
\be
	\lie{\J} (i^*G) \equiv 0.
\ee
Therefore, $\P$ may also be described as the {\em Killing
reduction}\cite{geroch} of $S$ with respect to the Killing field $\J$.

A manifold which arises in this way is always equipped with a
Riemannian metric of its own.%
\footnote{One must require that the integral curves of the Killing 
vector field do not come arbitrarily close to one another. This condition
is satisfied in our case.}
Although $i^*G$ is ``constant'' on the integral curves of $\J$, it is
not degenerate in that direction.  However, by subtracting off the
component in the direction of $\J$,
\be \label{g_defined}
\tilde{g} := \bigg[ G - \frac{1}{2\hbar}
(\Psi\otimes\Psi + \J\otimes\J ) \bigg]\bigg|_S,
\ee
we obtain a symmetric tensor field which agrees with $i^*G$ when
acting on vectors orthogonal to $\J$, is constant along $\J$ and is
degenerate only in the direction of $\J$.  Therefore $\tilde{g}$
defines a {\it strongly non-degenerate Riemannian metric} $g$ on $\P$.
It is a simple matter to verify that $g$, when combined with the
symplectic structure $\w$, equips the quantum phase space with the
structure of a {\em K\"ahler manifold}.

\subsubsection{Quantum observables}

According to Def. ~\ref{defn_observable_fn}, quantum mechanical
observables may be represented by real-valued functions on the quantum
phase space.  Unfortunately, these functions have still been defined
in terms of self-adjoint operators on the Hilbert space.  Our goal,
however, is a formulation of quantum mechanics which is intrinsic to
the projective space; we wish to avoid any explicit reference to the
underlying Hilbert space.  We now explain how this deficiency may be
overcome.

Since the Schr\"odinger vector field $Y_{\hat{F}} = \hvf{F}$ generates
a one-parameter family of unitary transformations on $\H$, $\hvf{F}$
preserves not only the symplectic structure $\W$, but the metric $G$
as well; $\hvf{F}$ is also a {\em Killing vector field}.  This fact
also holds for the corresponding observable function $f$; the
Hamiltonian vector field $X_f$ associated to any observable function
$f$ is also a Killing vector field on $(\P, g)$.  We will see that it
is this property which characterizes the set of observable functions
on the quantum phase space.

Let us begin by recalling a general property of Killing vector fields.
Since the calculations are somewhat involved, we will now use
Penrose's abstract index notation (which, as already pointed out, is
meaningful also on infinite-dimensional Hilbert manifolds.)  Let
$X^\a$ be any Killing vector field on $\P$.  Then, by definition,
$\grad_\a X_\b + \grad_\b X_\a = 0$, where $\grad$ denotes the
(Levi-Civit\`a) derivative operator associated to the metric $g$.
Therefore, $K_{\a\b} := \grad_\a X_\b$ is necessarily skew-symmetric.
As one can easily verify,\cite{wald} $K$ satisfies the identity%
\be
\grad_\a K_{\b\c} = R_{\c\b\a}{}^\de X_\de.
\ee
(Our conventions are such that for any 1-form $k_\c$,
$R_{\a\b\c}{}^\delta k_\delta = (\grad_\a \grad_\b - \grad_\b
\grad_\a) k_\c$.)  As a consequence, the Killing vector field $X$ is
{\em completely determined} by its value and first covariant
derivative at a single point.  (See \cite{ashtekar} or Appendix C of
Ref.~\cite{wald} for a discussion of this useful fact.)

Now suppose that the above Killing vector field is generated by the
observable function $f$. (Below, it will be understood that $X =
\hvf{f}$.) Since
\be
\w_\a{}^\c K_{\c\b} = - \w_\a{}^\c \grad_\b X_\c
		    = \grad_\a \grad_\b f,
\ee
$K$ satisfies the additional property that $\w_\a{}^\c K_{\c\b}$ is
symmetric.  (It then defines a bounded, skew-self-adjoint operator on
each tangent space.)  By considering the coupled differential
equations: 
\ba \grad_\a f = \w_{\c\a} X^\c, \\ \grad_\a X_\b =
K_{\a\b}, \\ \grad_\a K_{\b\c} = R_{\c\b\a}{}^\de X_\de, \ea 
we see \cite{thesis} that {\em any observable function is completely
determined by its value and first two derivatives at a single point!}
This fact motivates
\begin{definition}\label{Sp}
For each point $p\in\P$, let ${\cal S}_p$ consist of all triples,
$(\lambda, X_\a, K_{\a\b})$, where $\lambda$ is a real number, $X_\a$
is a covector at $p$, and $K_{\a\b}$ is a 2-form at $p$ for which
 $\w_\a{}^\c K_{\c\b} = \w_\b{}^\c K_{\c\a}$.  We call ${\cal S}_p$
the {\em algebra of symmetry data at $p$}.
\end{definition}
\noindent Thus, any observable function $f$ determines an element
$(\lambda, X, K) \in {\cal S}_p$ ($\lambda$ provides the value of $f$
at $p$), and $f$ is completely determined by this symmetry data.  The
algebra of quantum observables is then isomorphic to a subset of
${\cal S}_p$.  The converse to this statement is provided by
\cite{thesis}:
\begin{theorem}\label{thm_data}
For any element $(\lambda, X, K)$ of ${\cal S}_p$, there exists an
observable function $f : \P \rightarrow \R$ such that
$f(p) = \lambda$,  $(\hvf{f})_\a = X_\a$
and $\grad_\a (\hvf{f})_\b = K_{\a\b}$.
\end{theorem}
Therefore, the space of observable functions is isomorphic to the
algebra of symmetry data ${\cal S}_p$ for any $p\in\P$.  We will
utilize this result in section \ref{sec3.B}.  For the purposes of this
section, however, the main use of the above theorem is the following.
Recall that the Hamiltonian vector field of any observable function is
a Killing vector field.  Conversely, let $f$ be any smooth,
real-valued function on $\P$ for which $\hvf{f}$ is also a Killing
vector field.  Of course, the value and first two covariant
derivatives of $f$ at $p\in\P$ determine an element of ${\cal S}_p$.
Therefore, by Thm.~\ref{thm_data}, $f$ is an observable function.
This important result is expressed in
\begin{cor}
A smooth function $f:\P\rightarrow\R$ is observable if and only if its
Hamiltonian vector field is also Killing.
\end{cor}
Therefore, as in classical mechanics, the space of quantum observables
is isomorphic to the space of smooth functions on the phase space
whose Hamiltonian vector fields are infinitesimal symmetries of the
available structure.  However, now the available structure includes
not just the symplectic structure but also the metric. Hence, unlike
in classical mechanics, this function space is an extremely small
subset of the set of all smooth functions on $\P$.  This should not be
terribly surprising; as an example consider the finite-dimensional
case, for which the function space is infinite-dimensional while the
algebra of Hermitian operators is finite-dimensional.

\subsubsection{Quantum uncertainty}

In analogy with our considerations of the symplectic geometry of $\P$,
let us consider the Riemann bracket defined by the quantum metric $g$.
We denote the $g$-Riemann bracket of $f,k : \P \rightarrow \R$ by
\be
( f, k ) := \frac{\hbar}{2} g( \hvf{f}, \hvf{k} )
	  = \frac{\hbar}{2} (\grad_\a f) g^{\a\b} (\grad_\b k).
\ee
If $f$ and $k$ correspond to the expectation value functions
$F, K : \H \rightarrow \R$, then, by \eqn{g_defined},
\be
\{ F, K \}_+  = \frac{\hbar}{2} G( \hvf{F}, \hvf{K} )
= \frac{\hbar}{2} \tilde{g} ( \hvf{F}, \hvf{K})
+ \frac{1}{4} G( \J,\hvf{F} ) G( \J,\hvf{K} )
= \pi^* \big[ (f,k) + fk \big],
\ee
where we have used the fact that for any observable $\hat{F}$,
\be
G(\J, \hvf{F})|_\Psi = - G(J\Psi, \hvf{F}) 
= \W( \hvf{F}, \Psi ) = \Psi \circ (\d F) = 2F.
\ee

Therefore, the observable function which corresponds to the Jordan
product of $\hat{F}$ and $\hat{K}$ is given not by the $g$-Riemann
bracket of $f$ and $k$, but by the quantity
\be \label{symmetric_bracket2}
\{ f, k \}_+  := ( f, k) + fk,		
\ee
where we have utilized a minor notational abuse to emphasize the
relationship with \eqn{symmetric_bracket}.  The above quantity will be
called the {\em symmetric bracket} of $f$ and $k$.

Note that the $g$-Riemann bracket of two observables, while not
necessarily an observable itself, is a physically meaningful quantity;
it is simply the quantum covariance function (see the comment
following \eqn{new_unc_reln}).  In particular, $(f, f)(p)$ is exactly
the squared uncertainty of $f$ at the quantum state labeled by $p$;
\be \label{new_uncertainty}
(\Delta f)^2(p) := (\Delta \hat{F})^2(\pi^{-1}p) = (f, f)(p).
\ee

In order to obtain a feeling for the physical meaning of the quantum
covariance, notice that the uncertainty relation of \eqn{new_unc_reln}
assumes the form
\be
(\Delta f) (\Delta k) \ge
\bigg( \frac{\hbar}{2} \{ f, k\}_\w \bigg)^2 + ( f, k )^2.
\ee
As one can easily show, the standard uncertainty relation
\be
(\Delta f) (\Delta k) \ge
\bigg( \frac{\hbar}{2} \{ f, k\}_\w \bigg)^2
\ee
is saturated at the state $p \in \P$ if and only if $(f, k)(p) = 0$
{\em and} $\hvf{f} \propto j(\hvf{k})$, where $j$ is the complex
structure on $\P$ (compatible with $\omega$ and $g$).  The quantum
covariance $(f, k)(p)$ therefore measures the ``coherence'' of the
state $p$ with respect to the observables $f$ and $k$.

Let us conclude this sub-section by reproducing a result of Anandan
and Aharanov\cite{anandan}. Suppose that the Hamiltonian operator of a
quantum system is bounded and let $h$ be the corresponding observable
function.  By definition of the Riemann bracket, the uncertainty of
$h$ is given by $(\Delta h)^2 = \frac{\hbar}{2} g(\hvf{h}, \hvf{h})$.
Therefore, apart from the constant coefficient, the uncertainty in the
energy is exactly the length of the Hamiltonian vector field which
generates the time-evolution. Thus, the energy uncertainty can be
thought of as `the speed with which the system moves through the
quantum phase space'; during its evolution, the system passes quickly
through regions where the energy uncertainty is large and spends more
time in states where it is small.

\subsubsection{The measurement process}

Some of the most significant aspects of quantum mechanics involve the
measurement process and, without a complete geometric description of
these issues, our program would be incomplete. In this sub-section we
will sketch the desired geometric description, including the case when
the spectrum of the operator is continuous. A more complete discussion
may be found in \cite{thesis}.

Let $\Psi_0 \in S$ be an arbitrary normalized element of the Hilbert
space, and let $p_0 = \pi(\Psi_0)$ be its projection to $\P$. Of
obvious interest, in the context of measurement, is the function
$\tilde{\delta}_{\Psi_0} :\Psi\in S \mapsto \big| \< \Psi_0 , \Psi \>
\big|^2$.  Since $\tilde{\delta}_{\Psi_0}$ is independent of the phase of
$\Psi$ it defines a function $\delta_{p_0}$ on $\P$ via
\be
\delta_{p_0}(p) := \tilde{\delta}_{\Psi_0} ( \pi^{-1}p )
= \big| \< \pi^{-1}(p_0),  \pi^{-1}(p) \> \big|^2.
\ee
If the quantum system is in the state labeled by $p_0$ when a
measurement is performed, the relevant quantum mechanical probability
distribution is determined by the function $\delta_{p_0}$.  We
therefore desire a description of this function which does not rely
explicitly on the underlying Hilbert space.  This is provided by
\begin{theorem}\label{thm_geodesics}
Given arbitrary points $p_0, p \in \P$ there exists a (closed)
geodesic which passes through $p_0$ and $p$. Further, $d_{p_0}(p) =
\cos^2\left( \sigma(p_0, p) / \sqrt{2\hbar} \right)$, where
$\sigma(p_0, p)$ denotes the geodesic separation of $p_0$ and $p$.
\end{theorem}

A few comments regarding Thm.~\ref{thm_geodesics} are in order.
First, if $\pi^{-1}(p_0)$ and $\pi^{-1}(p)$ are non-orthogonal, then
the above-mentioned geodesic is unique, up to re-parameterization.
Next, since all geodesics on $\P$ are {\em closed}, the geodesic
distance $\sigma(p_0, p)$ is, strictly speaking, ill-defined.  Due to
the periodicity of the cosine function, however, the ``transition
amplitude function'', $d_{p_0}$, is insensitive to this ambiguity.
For the sake of precision, by $\sigma(p_0, p)$, we will mean the {\em
minimal} geodesic distance separating $p_0$ and $p$.

Suppose one is dealing with the measurement of an operator $\hat{F}$
with discrete, non-degenerate spectrum, and let $f$ be the
corresponding observable function.  Each eigenspace of $\hat{F}$ is
one complex-dimensional, and therefore determines a single point of
 $\P$.  Denote these eigenstates by $p_i$.  Suppose the system is in
the state labeled by the point $p_0$ when an ideal measurement of $f$
is performed.  We know that the system will `collapse' to one of the
states $p_i$.  Theorem~\ref{thm_geodesics} provides the corresponding
probabilities.  It is interesting to notice that the probability of
collapse to an eigenstate $p_i$ is a monotonically decreasing function
of the geodesic separation of $p_0$ and $p_i$; the system is more
likely to collapse to a nearby state than a distant one.

We now have a description of the probabilities associated with the
measurement process, but there are two deficiencies to be remedied.
The eigenstates $p_i$ above have been defined in terms of the
algebraic properties of the operator $\hat{F}$.  For our program to be
complete, we require a definition of these eigenstates which does not
refer to the Hilbert space explicitly.  Next, the above discussion was
limited to the measurement of an observable with discrete,
non-degenerate spectrum.  We will describe the generic situation in
two steps.  First, we consider the measurement of observables with
discrete, but possibly degenerate, spectra.  This will require the
aforementioned description of the eigenstates.  We will then be
prepared to consider measurement of observables with continuous
spectra.

Let us first examine the notions of eigenstates and eigenvalues of an
observable operator $\hat{F}$. Let $F : \H \rightarrow \R$ and $f :
\P \rightarrow \R$ be the expectation value and corresponding
observable function, respectively.  A vector $\Psi\in\H$ is an
eigenstate of $\hat{F}$ iff $\hat{F}\Psi = \lambda \Psi$, for some
(real) $\lambda$.  Alternatively, by \eqn{schrodinger_vf},
\be
\hvf{F}|_\Psi = Y_{\hat{F}}|_\Psi = (\lambda / \hbar) \J|_\Psi.
\ee
That is, $\Psi$ is an eigenstate of $\hat{F}$ if and only if the
Hamiltonian vector field, $\hvf{F}$ is {\em vertical} (i.e., purely
gauge) at $\Psi$.  This will be the case if and only if $\hvf{f}$
vanishes at $\pi(\Psi)$;  $\pi(\Psi)$ is then a critical point of the
function $f$.  Evidently, the corresponding eigenvalue is exactly the
(critical) value of $f$ at $\pi(\Psi)$.  In summary:
\begin{definition}
Let $f:\P\rightarrow\R$ be an observable function.  Critical point of
$f$ are called {\em eigenstates} of $f$.  The corresponding critical
values are called {\em eigenvalues}.
\end{definition}

We now consider the measurement of an observable $f$ whose spectrum is
discrete but possibly degenerate (of course, the ``spectrum of $f$''
coincides, by definition, with the spectrum of the corresponding
operator $\hat{F}$).  Let $\lambda$ be a degenerate eigenvalue of
$\hat{F}$, and denote by $\tilde{\cal E}_\lambda$ the associated
eigenspace of $\H$.  Associated to this eigenspace is a sub-manifold,
${\cal E}_\lambda$, of $\P$, which we shall call the {\em
eigenmanifold} associated to $\lambda$.  Suppose that the system is
prepared in the state labeled by $p_0$ and let $\Psi_0 \in
\pi^{-1}p_0$.  The postulates of ordinary quantum mechanics assert
that measurement of $f$ will yield the value $\lambda$ with
probability $\< {\Bbb P}_\lambda \Psi_0, {\Bbb P}_\lambda \Psi_0 \> =
\< \Psi_0, {\Bbb P}_\lambda \Psi_0 \> = \left| \left< \Psi_0, {\Bbb
P}_\lambda \Psi_0 / \| {\Bbb P}_\lambda \Psi_0 \| \right> \right|^2$,
where ${\Bbb P}_\lambda$ is the projector onto the relevant eigenspace
of $\hat{F}$. {}From the above considerations, we know that this
probability may be expressed in terms of the geodesic separation of
the points $p_0$ and $\pi\left( {\Bbb P}_\lambda \Psi_0 / \| {\Bbb
P}_\lambda \Psi_0 \|\right) \in {\cal E}_\lambda$.  We will denote the
latter point by ${\Bbb P}_\lambda(p_0)$.

What sets the point ${\Bbb P}_\lambda(p_0)$ apart from all other
elements of  ${\cal E}_\lambda$?  We need only notice that for any
point  $\Phi \in \tilde{\cal E}_\lambda \cap S$,
\be\label{projection-ineq}
\left| \< \Psi_0, \Phi \> \right|^2 =
\left| \< \Psi_0, {\Bbb P}_\lambda \Phi \> \right|^2 =
\left| \< {\Bbb P}_\lambda \Psi_0, \Phi \> \right|^2 \le
\| {\Bbb P}_\lambda \Psi_0 \|^2.
\ee
Therefore, of all elements $\Phi \in \tilde{\cal E}_\lambda$ with unit
normalization, that which maximizes the quantity $|\<
\Psi_0,\Phi\>|^2$ is simply ${\Bbb P}_\lambda \Psi_0 / \| {\Bbb
P}_\lambda \Psi_0 \|$, i.e., that to which the state $\Psi_0$ will
`collapse' in the event that measurement of $\hat{F}$ yields the value
$\lambda$.  Therefore, by Thm.~\ref{thm_geodesics}, ${\Bbb
P}_\lambda(p_0)$ is simply that point of ${\cal E}_\lambda$ which is
{\em nearest} $p_0$!

We may now describe the measurement of an observable with discrete
spectrum as follows.  Suppose that immediately prior to measurement of
$f$, the system is in the state $p_0 \in \P$.  Denote by $\lambda_i$
the critical values of $f$ and, by ${\cal E}_{\lambda_i}$ the
corresponding eigenmanifolds.  Interaction with the measurement device
causes the system to be projected to one of the eigenmanifolds, say
${\cal E}_{\lambda_0}$.  ``Realizing that it collapsed'' to the state
${\Bbb P}_{\lambda_0}(p_0)$, the system returns what is knows to be
the value of the observable under consideration, i.e., $f\left( {\Bbb
P}_{\lambda_0}(p_0)\right) = \lambda_0$.  Of course, the probability
that measurement causes reduction to ${\cal E}_{\lambda_0}$ is given
by $\cos^2 \left( \sigma(p_0,{\cal E}_{\lambda_0}) / \sqrt{2\hbar}
\right)$, where, $\sigma(p_0,{\cal E}_{\lambda_0})$ denotes the {\em
minimal} geodesic separation of $p_0$ and the sub-manifold ${\cal
E}_{\lambda_0}$.

Now let us study the generic case.  Let $\hat{F}$ be any observable
operator on $\H$, the spectrum of which is allowed to be continuous.
We first need a definition of the spectrum of $\hat{F}$ in terms of
the corresponding observable function $f:\P\rightarrow\R$.  Recall the
standard definition \cite{reed-simon}: $\lambda$ is an element of the
spectrum ${\mathrm sp}(\hat{F})$ if and only if the operator $\hat{F}
- \lambda\hat{1}$ is {\em not} invertible.  Equivalently, $\lambda \in
{\mathrm sp}(\hat{F})$ iff given any positive $\varepsilon\in\R$,
$\exists \Psi\in S$ such that $\| \hat{F}\Psi - \lambda\Psi \| <
\varepsilon$.  This condition guarantees that, to arbitrary precision,
$\lambda$ is an approximate eigenvalue of $\hat{F}$.

Using Eqs.~(\ref{symmetric_bracket2}) and (\ref{new_uncertainty}), we
may write the (square of the) above quantity as
\be
\| \hat{F}\Psi - \lambda \Psi \|^2 =
\left. \{ f - \lambda, f - \lambda \}_+ \right|_{\pi(\Psi)} =
\left.\left[ (\Delta f)^2 + (f - \lambda)^2
\right] \right|_{\pi(\Psi)}.
\ee
This equation allows us to define the spectrum of $\hat{F}$ in terms
of the function $f:\P\rightarrow\R$;
\begin{definition}
The {\em spectrum} ${\mathrm sp}(f)$ of an observable $f$ consists of
all real numbers $\lambda$ for which the function $n_{\lambda} : \P
\rightarrow \R \cup \{\infty\}, \;\; n_{\lambda} : p \mapsto \left[
\big(\Delta f\big)^2(p) + \big(f(p) - \lambda\big)^2 \right]^{-1}$ is
{\em unbounded}.
\end{definition}
\noindent
Of course, a point at which $n_\lambda = \infty$ corresponds to an
eigenstate of $f$.

The next step is a description of the spectral projection operators.
Let $\Lambda$ be a closed subset of the spectrum ${\mathrm sp}(f)$ of
$f$, and denote by ${\Bbb P}_{\hat{F}, \Lambda}$ the projection
operator associated to $\hat{F}$ and $\Lambda$. In analogy with the
above, put $\tilde{\cal E}_{\hat{F}, \Lambda} = \big\{ {\Bbb
P}_{\hat{F}, \Lambda}\Psi \, \big| \, \Psi\in\H - \{{\mathbf0}\}
\big\}$, and let ${\cal E}_{f, \Lambda}$ denote the projection of
$\tilde{\cal E}_{\hat{F}, \Lambda}$ to $\P$.  Note that the set
$\tilde{\cal E}_{\hat{F}, \Lambda}$---the analogue of the eigenspace
above--- actually {\em is} the eigenspace of ${\Bbb P}_{\hat{F},
\Lambda}$ corresponding to the eigenvalue $1$.  Therefore, we have
${\cal E}_{f, \Lambda}$ consists of the critical points of an
expectation value function, associated to the critical value $1$.
Unfortunately, this expectation value function is not directly
expressible in terms of $f$ and $\Lambda$.  We must look for an
alternative description of the sub-manifold ${\cal E}_{f, \Lambda}$.

In the representation defined by the operator $\hat{F}$, elements of
$\tilde{\cal E}_{\hat{F}, \Lambda}$ have support on $\Lambda$.
Therefore, $\Psi \in \tilde{\cal E}_{\hat{F}, \Lambda}$ iff $\<
\hat{F} \>_\Psi \in \Lambda^n \; \forall n>0$, where $\Lambda^n$
denotes the image of $\Lambda$ under the map $\lambda \mapsto
\lambda^n$.  Recall that $\{f, f\}_+$ is the (projection to $\P$ of
the) expectation value of $\hat{F}^2$.  In general, the expectation
value of $\hat{F}^n$ projects to the $n$-fold symmetric product $\{f,
\{f, \{f, \ldots\}_+ \}_+ \}_+$.  Therefore, we have
\be\label{eq-E-defined}
{\cal E}_{f, \Lambda} = \{ q\in\P \big| 
	\{f, \{f, \{f,\ldots\}_+ \}_+ \}_+\big|_q
	\in \Lambda^n \; \forall n>0 \},
\ee
where there are $n$ factors of $f$ occurring above.

Having obtained a description of ${\cal E}_{f, \Lambda}$, we may now
define the spectral projections in a manner intrinsic to the
projective space.  By precisely the same reasoning surrounding
\eqn{projection-ineq}, ${\Bbb P}_{f, \Lambda}$ maps a point $p\in\P$
to that element of ${\cal E}_{f, \Lambda}$ which is nearest $p$.

The measurement process may then be described as follows.  Suppose the
quantum system is in the state labeled by the point $p_0\in\P$ at the
instant an experimenter performs a measurement of the observable $f$.
Following the rules of quantum mechanics, she ``asks the system''
whether the value of $f$ lies in $\Lambda$---a closed subset of
${\mathrm sp}(f)$, which she is free to choose.  The experimental
apparatus drives the system to one of two states---either ${\Bbb
P}_{f, \Lambda}(p_0)$ or ${\Bbb P}_{f, \Lambda^c}(p_0)$, where
$\Lambda^c$ is the (closure of the) complement, in ${\mathrm sp}(f)$,
of $\Lambda$.  The system is reduced to the former with probability
\[
 \cos^2 \left(
 \frac{\sigma ( p_0, {\Bbb P}_{f, \Lambda}(p_0))}{\sqrt{2\hbar}}
 \right);
\]
in this event, the experiment yields the positive result
($f\in\Lambda$).  Having precisely prepared the system in the state
$p_0$, the experimenter may then infer the value $f({\Bbb P}_{f,
\Lambda}(p_0))$ of the observable $f$.  The probability of reduction
to the latter state is obtained by replacing $\Lambda$ by $\Lambda^c$
above.

Note that this description encompasses all measurement situations.  In
the event that the spectrum of $f$ is discrete and non-degenerate, the
experimenter may choose to let $\Lambda_i$ contain the single
eigenvalue $\lambda_i$.  Moreover, she may measure all of the
projections simultaneously.  In this way, one recovers the first
familiar description of the measurement process.  Note also that,
while the above discussion of the spectral projections may seem
complicated and somewhat unnatural at first, the definition of the
spectral projection operators on the Hilbert space has the same
features. (Indeed, most text books simply skip this technical
discussion.) This is simply one of the technical complications that
the geometric formalism inherits from the Hilbert space framework.

To conclude, we wish to emphasize that the topic of our discussion has
been ordinary quantum mechanics.  We have just restated the well-known
quantum mechanical formalism in a language intrinsic to the true space
of states---the quantum phase space, $\P$; no new ingredients have
been added to the physics.  A particularly attractive feature of the
formalism, however, is the fact that slight modifications of the
standard picture naturally present themselves.  For example, using
many of these geometric ideas, Hughston\cite{hughston2} has explored a
novel approach to the measurement problem in terms of stochastic
evolution.

\subsection{The postulates of quantum mechanics} \label{sec2.D}

Let us collect the results obtained in the first three sub-sections.

We have formulated ordinary quantum mechanics in a language which is
intrinsic to the true space of quantum states---the projective Hilbert
space $\P$.  As in classical mechanics, observables are smooth,
real-valued functions which preserve the kinematic structure.  Being a
K\"ahler manifold, $\P$ is a symplectic manifold.  The role of the
quantum symplectic structure is precisely that of classical mechanics;
it defines both the Lie algebraic structure on the space of
observables and generates motions including the time-evolution.

There are, however, two important features of quantum mechanics which
are not shared by the classical description.  First, the phase space
is of a very particular nature; it is a K\"ahler manifold and, as we
will see in section \ref{sec3.B}, one of a rather special
type---namely one of constant holomorphic sectional curvature.%
\footnote{In this sense, the quantum framework is actually a special 
case of the classical one!}
The second difference lies in the probabilistic aspects of the
formalism, which is itself intimately related to the presence of the
Riemannian metric. More generally, this metric describes those quantum
mechanical features which are absent in the classical theory---namely,
the notions of uncertainty and state reduction.  For example, the
transition probabilities which arise in quantum mechanics are
determined by a simple function of the geodesic distance between
points of the phase space.

These results are most easily summarized by stating the postulates in
the geometric language:
\begin{itemize}
\item[(${\cal H}$)] {\it Physical states: } Physical states of the
quantum system are in one-to-one correspondence with points of a
K\"ahler manifold ${\cal P}$, which is a projective Hilbert space.%
\footnote{We will see in section \ref{sec3.B} that quantum phase
spaces $\P$ can be alternatively singled out as K\"ahler manifolds
which admit maximal symmetries.  This provides an intrinsic
characterization without any reference to Hilbert spaces.}
\item[(${\cal U}$)] {\it K\"ahler evolution: } The evolution of the
system is determined by a flow on $\P$, which preserves the K\"ahler
structure. The generator of this flow is a densely defined vector
field on $\P$.
\item[(${\cal O}$)] {\it Observables: } Physical observables are
represented by real-valued, smooth functions $f$ on $\P$ whose
Hamiltonian vector fields $X_f$ preserve the K\"ahler structure.
\item[(${\cal P}$)] {\it Probabilistic interpretation: } Let $\Lambda
\subset \R$ be a closed subset of the spectrum of an observable $f$,
and suppose the system is in the state corresponding to the point $p
\in \P$.  The probability that measurement of $f$ will yield an
element of $\Lambda$ is given by
\begin{equation}
\delta_p(\Lambda) = \cos^2\left(\frac {\sigma(p, \;{\Bbb P}_{f,\Lambda}
(p))} {\sqrt{2\hbar}} \right),
\end{equation}
where ${\Bbb P}_{f,\Lambda}(p)$ is the point, closest to $p$, in the
space ${\cal E}_{f, \Lambda}$, defined bye \eqn{eq-E-defined}.
\item[(${\cal R}_D$)] {\it Reduction, discrete spectrum: } Suppose the
spectrum of an observable $f$ is discrete.  This spectrum provides the
set of possible outcomes of the ideal measurement of $f$.  If
measurement of $f$ yields the eigenvalue $\lambda$, the state of the
system immediately after the measurement is given by the associated
projection, ${\Bbb P}_{f,\lambda}(p)$, of the initial state $p$.
\item[(${\cal R}_C$)] {\it Reduction, continuous spectrum: } A closed
subset $\Lambda$ of the spectrum of $f$ determines an ideal
measurement that may be performed on the system.  This measurement
corresponds to inquiring whether the value of $f$ lies in $\Lambda$.
Immediately after this measurement, the state of the system is given
by ${\Bbb P}_{f,\Lambda}(p)$ or ${\Bbb P}_{f,\Lambda^c}(p)$, depending
on whether the result of the measurement is positive or negative,
respectively.
\end{itemize}
In the last postulate, $\Lambda^c$ is the closure of the complement,
in the spectrum of $f$ of the set $\Lambda$.  Although the first
``reduction postulate''  is a special case of the second, both have
been included for comparison with standard textbook presentations.

To conclude, although it is not obvious from textbook presentations,
the postulates of quantum mechanics can be formulated in an
intrinsically geometric fashion, without any reference to the Hilbert
space. The Hilbert space and associated algebraic machinery provides
convenient technical tools. But they are not
essential. Mathematically, the situation is similar to the discussion
of manifolds of constant curvature.  In practice, one often
establishes their properties by first embedding them in $\R^n$
(equipped with a flat metric of appropriate signature). However, the
embedding is only for convenience; the object of interest is the
manifold itself. There is also a potential analogy from physics,
alluded to in the Introduction.  Perhaps the habitual linear
structures of quantum mechanics are analogous to the inertial rest
frames in special relativity and the geometric description summarized
here, analogous to Minkowski's reformulation of special relativity.
Minkowski's description paved the way to general relativity.  Could
the geometric formulation of quantum mechanics lead to a more complete
theory one day?

\section{A unified framework for generalizations of quantum 
mechanics} \label{sec3}

There are three basic elements of the quantum mechanical formalism
which may be considered for generalization: the state space, the
algebra of observables and the dynamics.  The framework developed in
section \ref{sec2} suggests avenues for each of these.  First, while
it is not obvious how one might ``wiggle'' a Hilbert space, one may
generalize the quantum phase space by considering, say, the class of
all K\"ahler manifolds $\{ \M, g, \w \}$.  The geometric language also
suggests an obvious generalization of the space of quantum
observables: one might consider the space of {\em all} smooth,
real-valued functions on the phase space.  Finally, whether or not one
chooses either of these extensions, one may consider generalized
dynamics which, as in classical mechanics, preserves only the
symplectic structure. Thus, one might require the dynamical flow to
preserve only the symplectic structure, and not necessarily the
metric.

While each of these structures may be extended separately, they are
intimately related and construction of a complete, consistent
framework is a highly non-trivial task. Thus, for example, if one
allows {\it all} K\"ahler manifolds as possible quantum phase spaces,
is seems very difficult to obtain consistent probabilistic predictions
for outcomes of measurements. More generally, the problem of
systematically analyzing viable, non-trivial generalizations of the
the kinematic structure would be a major undertaking, although the
pay-off may well be exceptional. Modification of dynamics, on the
other hand, is easier at least in principle. Therefore, we will first
consider these in section \ref{sec3.A} and return to kinematics in
section \ref{sec3.B}.

\subsection{Generalized dynamics} \label{sec3.A}

Let us then suppose that we continue to use a projective Hilbert space
for the quantum phase space, and let dynamics be generated by a
preferred Hamiltonian function. However, let us only require that
time-evolution should preserve the symplectic structure $\omega$ (as
in classical mechanics), and not necessarily the metric $g$. {}From
the viewpoint of the geometric formulation, this is the simplest and
most obvious generalization of the standard quantum dynamics.%
\footnote{Recall that the Hamiltonian need not be an observable
function in the sense of section \ref{sec2.B}. We could extend the
kinematical set up as well and regard any smooth function on $\P$ as
an observable function.  (This would be analogous to Weinberg's
\cite{weinberg} proposal which, however, was made at the level of the
Hilbert space $\H$ rather than the quantum phase space $\P$.) We have
refrained from doing this because the required extension of
measurement theory is far from obvious.}  
The idea is reminiscent of the ``non-linear Schr\"odinger equations''
that have been considered in the past. Therefore, it is
natural to ask if these there is a relation between the two.  We will
see that the answer is in the affirmative. Furthermore, the geometric
framework provides a unified treatment of these proposals and makes
the relation between them transparent, thereby enabling one to correct
a misconception.

Let us begin by defining the the class $\ch$ of Hamiltonians we now
wish to consider. $\ch$ will consist of densely defined functions $f$
on $\P$ satisfying the following properties: i) $f$ is smooth on its
domain of definition; and ii) the Hamiltonian vector field it defines
generates a flow on all of $\P$.  In particular, The Hamiltonian
functions we considered in section \ref{sec2}---expectation values of
a possibly unbounded self-adjoint operator---belong to $\ch$ but they
constitute only a `small subset' of $\ch$.

The existing proposals of non-linear dynamics refer to flows in the
full Hilbert space $\H$ rather than in the quantum phase space $\P$.
To compare the two, we need to lift our flows to $\H$.  Let us begin
by recalling that it is natural to regard $\P$ as a reduced phase
space, resulting from the first class constraint $ C(\Psi) := \< \Psi,
\Psi\> -1 =0$ on $\H$ (see Eq.~\ref{constraint}).  Therefore, a
function $f$ on $\P$ admits a natural lift $F$ to $S$, the unit sphere
in $\H$. This function $F$ on $S$ is constant along the integral
curves of the vector field $\J$ which generates phase rotations;
$\lie{\J}f = 0$. Denote the space of these lifts by $\tilde\ch$.

To discuss dynamics on $\H$, we need to extend these functions%
\footnote{Strictly speaking, we only need to extend the dynamical flow
off $S$. However, to compare our results with Weinberg's \cite{weinberg}
we need to consider extensions of Hamiltonians. In discussions on
generalized dynamics \cite{weinberg,birula}, the issue of domains of
definition of operators is generally ignored. Our treatment will be at
the same level of rigor. In particular, we will ignore the fact that
our Hamiltonian functions and vector fields are only densely
defined.}
off $S$. {}From the reduced phase space viewpoint, the extension is
completely arbitrary. For, we can construct the Hamiltonian vector
field on $\H$ generated by {\it any} extension $F_{ext}$.  The
restriction to $S$ of this vector field does depend on the extension
but the the {\it horizontal part} of the restriction---i.e., the part
orthogonal to $\J$---does not. Hence, the projection of the vector
field to $\P$ agrees with $X_f$, irrespective of the choice of the
initial extension. Thus, the generalized dynamics generated by a given
Hamiltonian function $f$ on $\P$ can be lifted to a whole family of
flows on $\H$, all of which, however, evolve the physical quantum
states---elements of $\P$---in the same way. Because of this,
apparently distinct proposals for non-linear evolutions on $\H$ can in
fact be physically equivalent. This point is rather trivial from the
viewpoint of geometric quantum mechanics. The reason for our
elaboration is that---as we will see below---it has not been
appreciated in the Hilbert space formulations.

While the extension off $S$ of elements of $\tilde\ch$ is completely
arbitrary, one can use the standard quantum mechanical framework to
select a specific rule. Consider, to begin with, a bounded,
self-adjoint operator $\hat{F}$ on $\H$, and let $F$ be the
restriction to $S$ of the corresponding expectation value function.
There is then an obvious extension of $F$ to all of $\H$: set
$F_{ext}(\Psi) := \<\Psi, \hat{F}\Psi\>$ (which we denoted by $F$ in
section \ref{sec2.A}). One can restate this rule as:
\be\label{f_ext} 
F_{ext}(\Psi) := \|\Psi\|^2 F( \Psi / \|\Psi\| ).  
\ee
The advantage is that this equation may now be used to extend {\em
any} element $F$ of $\tilde\ch$ to all of $\punctH = \H-\{{\mathbf
0}\}$. Note that, with this preferred extension, the Hamiltonian
vector field $\hvf{F_{ext}}$ is homogeneous of degree one on
$\punctH$:
\be \label{hom}
\hvf{F_{ext}}(c\Psi) = c \hvf{F_{ext}}(\Psi) \quad \forall c\in \C.  
\ee 
Hence, the flow on $\punctH$ which is generated by $F_{ext}$ is
homogeneous, but fails to be linear unless $F$ is the restriction to
$S$ of the expectation value function defined by a self-adjoint
operator.  Next, it is easy to verify that these $F_{ext}$ {\em
strongly} commute with the constraint function $C(\Psi)$.  Hence the
flow along $\hvf{F_{ext}}$ preserves the constraint.  Therefore, the
specific extension considered above has the property that the
corresponding flow preserves not only the symplectic structure
$\Omega$ {\it but also the norm} on $\punctH$. However, unless $f$ is
an observable function, it does not preserve the metric $G$.

Note that, even if we consider just these preferred extensions, the
set of possible Hamiltonian functions on $\punctH$ has been extended
quite dramatically.  To see this, consider the case when $\H$ is
finite-dimensional. Then, the class of Hamiltonian functions on $\H$
allowed in standard quantum mechanics forms a {\it finite}-dimensional
real vector space; it is just the space of expectation value functions
constructed from self-adjoint operators. The space $\tilde\ch$, on the
other hand, is {\it infinite}-dimensional since its elements are in
one-to-one correspondence with smooth functions on $\P$. And each
element of $\tilde\ch$ admits an unique extension to $\punctH$ via
Eq.~(\ref{f_ext}). It is natural to ask if one can do something
`in-between'. Can we impose more stringent requirements to select a
class of potential Hamiltonians which is larger than that of
observable functions of section \ref{sec2} but smaller than $\ch$?
For example, one might imagine looking for the class of functions on
$\punctH$ whose Hamiltonian flows preserve not just the norms but also
the inner-product. It turns out, however, that this class consists
precisely of the expectation value functions defined by self-adjoint
operators; there is no such thing as a non-linear unitary flow on $\H$
\cite{thesis}. Despite the magnitude of our generalization, it seems
to be the only available choice.

We are now ready to discuss the relation between this generalization
and those that have appeared in the literature. Note first that
Eq.~(\ref{f_ext}) implies that there is a one-to-one correspondence
between elements of $\ch$ and smooth functions on $\punctH$ which are
gauge-invariant (i.e. insensitive to phase) and homogeneous of degree
two. (If $\H$ is viewed as a vector space over complex numbers, this
corresponds to homogeneity of degree one in both $\Psi$ and
$\bar{\Psi}$.)  It turns out that this is precisely the class of
permissible Hamiltonians that Weinberg \cite{weinberg} was led to
consider while looking for a general framework for non-linear
generalizations of quantum mechanics. Let us therefore call functions
on $\punctH$ satisfying Eq.~(\ref{f_ext}) {\em Weinberg functions},
and denote the space of these functions by ${\cal O}_W$. Our
discussion shows that there is a one-to-one correspondence between
smooth functions on the projective Hilbert space $\P$ and Weinberg
functions on the punctured Hilbert space $\punctH$;the homogeneity
restriction simply serves to eliminate the freedom in the extension of
the function on $\punctH$.  Thus the extension of quantum dynamics
that is immediately suggested by the geometrical framework reproduces
key features of Weinberg's proposal.

There are, however, considerable differences in the motivations and
general viewpoints of the two treatments. In particular, Weinberg
works with the Hilbert space $\H$ (and, without explicitly saying so,
sometimes with $\punctH$). However, he does assume at the outset that
elements $\Psi$ and $c\Psi$ of $\H$ define the same physical state of
the quantum system for all {\it complex} numbers $c$. Thus, although
it is not explicitly stated, his space of physical states is also
$\P$.  Therefore, it is possible to translate his constructions to the
the geometric language. As we will see below, the geometric viewpoint
is often clarifying.

Next, let us consider two specific examples of non-linear dynamics
that have been considered in the literature.  Each of these involves
the non-relativistic mechanics of a point particle moving in $\R^n$
and the Hilbert space consists of square-integrable functions on
$\R^n$. Therefore, it will be useful to reinstate a complex notation
for the remainder of this sub-section.  In this notation, an element
of ${\cal O}_W$ is a real-valued function $F(\Psi, \bar{\Psi})$ of
both $\Psi$ and its conjugate, which is homogeneous of degree one in
each argument. The Hamiltonian vector field generated by such a
function corresponds to
\be 
\hvf{F}[\Psi](x) = \frac{1}{\i\hbar} \frac{\delta
F}{\delta \Psi^*} ( x ).  
\ee
The simplest example of this type is provided by is the so-called
``non-linear Schr\"odinger equation''.  This equation is given by
\be \label{nse}
\i\hbar \frac{\partial \Psi}{\partial t}(x,t) = 
(\hat{H}_0 \Psi)(x,t) + \epsilon | \Psi(x,t) |^2 \Psi(x,t),
\ee
where $\hat{H}_0 = \hat{P}^2/2m + \hat{V}$ is the standard Hamiltonian
operator describing a non-relativistic particle under the influence of
a conservative force with potential $\hat{V}$.  Note that the quantity
$| \Psi(x,t) | $ is the modulus of $\Psi(x,t)$, {\em not} the norm
$\|\Psi\|$ of the state-vector $\Psi$. It is easy to verify that
(\ref{nse}) induces a flow on $\P$. This flow is Hamiltonian and the
generating function is the projection to $\P$ of the function
$\<\Psi, H_0\Psi\> + H_\epsilon(\Psi)$ on $\H$, where,
\be
 H_\epsilon(\Psi) := \frac{\epsilon}{2}\int \d^n \! x
\left[ \Psi^*(x,t) \Psi(x,t) \right]^2.
\ee
Thus, Eq.~(\ref{nse}) is indeed a specific example of our generalized
dynamics.

Note, however, that the dynamical vector field $X$ on $\H$ defined
directly by the non-linear Schr\"odinger equation is given by
\be X[\Psi](x) = \hvf{H_0}[\Psi](x) + X_\epsilon[\Psi](x), 
\ee 
where $X_\epsilon[\Psi](x) = (\epsilon/\i\hbar) |\Psi(x,t)|^2
\Psi(x,t)$. Clearly, it is {\it not} homogeneous in the sense of
Eq. (\ref{hom}), and hence is not generated by a Weinberg function.
Therefore, Weinberg was led to state that the ``results obtained by
the mathematical studies of this equation are unfortunately of no
use'' to the generalization he considered. However, we just saw that
the non-linear Schr\"odinger equation does correspond to generalized
dynamics on $\P$ and is therefore of `Weinberg type'. Hence, the
statement in quotes is somewhat misleading.

Let us elaborate on this point. If we first focus on physical states,
what matters is just the projected flow on $\P$. This in turn is
completely determined by the restriction $H_\epsilon \big|_S$ of
$H_\epsilon$ to $S$. Therefore, we may feel free to ignore the
behavior of $H_\epsilon$ off of $S$.  Of course, to construct a vector
field on $S$ which projects to the relevant one on $\P$, we can extend
$H_\epsilon \big|_S$ {\em arbitrarily} and compute the associated
Hamiltonian vector field.  In particular, we may extend it in the way
which Weinberg would suggest:
\be H_\epsilon^\prime(\Psi) := \| \Psi \|^2 H_\epsilon 
\big|_S(\Psi / \|\Psi\|).  
\ee 
Thus, we have seen explicitly that the flow on $\P$ which is defined
by $H_\epsilon$ may also be described by a Weinberg function! The
emphasis on the true space of states $\P$ of the geometric treatment
clarifies this point which seems rather confusing at first from the 
Hilbert space perspective.

The non-linear Schr\"odinger equation is a fairly simple example since
the generating function $H_\epsilon$ is itself homogeneous (but of the
``wrong'' degree to be a Weinberg function).  Let us now consider an
example which is more sophisticated.

In an effort to address the problem of combining systems which are
subject to a non-linear equation of motion, Bialynicki-Birula and
Mycielski were led to a {\em logarithmic} equation of motion
\cite{birula}. They began with a general equation of motion of the
form
\be \label{gen_nonlin}
\i\hbar \frac{\partial \Psi}{\partial t}(x,t) =
(\hat{H}_0\Psi)(x,t) + \alpha( | \Psi(x,t) |^2 ) \Psi(x,t),
\ee
and showed that physical considerations, particularly the requirement
that $\alpha$-term should not introduce interactions between otherwise
non-interacting subsystems, imposes severe restrictions the functional
form of $\alpha$. The only possibility is to have: $\alpha(\rho) = -b
\ln(a^n\rho)$ for some constants $a$ and $b$. (For details see
\cite{birula}).  Choosing units%
\footnote{The constant $a$ is a length scale of no physical
significance, since it may be altered by addition of a constant to the
Hamiltonian operator.}
with respect to which $a=1$, the vector field along which the system
evolves may be written as 
\be X[\Psi](x) = \hvf{H_0}[\Psi](x) +X_1[\Psi](x), \ee 
where $X_1[\Psi](x) = -(b/\i\hbar) \ln(|\Psi|^2) \Psi(x)$.  Again, the
extra term may be seen to be Hamiltonian; $X_1 = \hvf{H_1}$, where 
\be
H_1(\Psi) = b \int \d^n \! x \Psi^*(x) \Psi(x) \big[ 1 - 
\ln(\Psi^*(x) \Psi(x) \big].  
\ee
Since $H_1$ is phase-invariant, this is also an example of our
generalized dynamics.

Again, we can carry out the procedure used for the non-linear
Schr\"odinger equation to see that the corresponding motion on the
projective space may be described by use of a Weinberg function.  It
is not difficult to show that the corresponding homogeneous function
is given by
\be
 H_1^\prime(\Psi) = H_1(\Psi) + b\|\Psi\|^2 \ln( \| \Psi \|^2 ).
\ee
Therefore, the logarithmic equation induces a flow on $\P$ which may
be described by a function of the Weinberg type.

In retrospect, essentially any generalized dynamics of the form
specified by Eq. (\ref{gen_nonlin}) may be written as a Hamiltonian
flow on $\punctH$ by ``integrating'' the function $F$.  The resulting
Hamiltonian function is not likely to satisfy Weinberg's homogeneity
condition.  However, one may restrict it to the unit sphere, then
extend this restriction as in \eqn{f_ext}.  The resulting function
will generate a Hamiltonian vector field which differs from the
original one in general, but agrees with it on the unit sphere.
Hence, both will generate the same flows on the projective space.
This may be done with {\em any} Hamiltonian function on $\punctH$
which preserves the unit sphere.

\subsection{Characterization of the standard quantum kinematics}
\label{sec3.B}					

Let us now turn to the issue of generalizations of the kinematic
framework itself. As noted above, these extensions may well turn out
to be profound. However, they also appear to be much more difficult to
carry out. Therefore, as a first step, it is natural to ask what sets
ordinary quantum mechanics apart from its natural generalizations.
The purpose of this sub-section is provide such a characterization of
the standard kinematic framework.

Let us consider an {\em arbitrary} K\"ahler manifold $\{ \M, g, \w
\}$, which is to represent the phase space of a (radically!)
generalized quantum theory. (As before, $\M$ will be assumed to be a
Hilbert manifold, and $g$ and $\w$ to be strongly non-degenerate.)
The question is then: Are there natural conditions which
will single out ordinary quantum mechanics from this class of
generalizations?  As we saw in section \ref{sec2}, the observables of
ordinary quantum mechanics are smooth functions on the phase space
whose Hamiltonian vector fields satisfy Killing's equation; we
therefore let the observables of the generalized framework consist of
all smooth functions on $\M$ whose Hamiltonian flows are
isometries. Denote this set as
\be
  {\cal O} := \{ f: \M \mapsto \R \; | \; \lie{\hvf{f}} g = 0 \}.
\ee
(Note that while we call elements of ${\cal O}$ {\em observables},
we do not claim to possess a complete and consistent formalism
incorporating measurement. Indeed, this seems an impossible feat at
the present level of generality.)

It should be emphasized that there are K\"ahler manifolds which do not
admit a single observable function (other than constants); the torus
is an example.  Let us amplify this point briefly.  Since the
discussion leading to Def.~\ref{Sp} is valid for an arbitrary K\"ahler
manifold, any observable is completely determined by its value and
first two derivatives at a single point.  As before, we will denote by
${\cal S}_p$ the set of symmetry data at the point $p \in \M $ (if
$\lambda$, $X$ and $K$ are 0-, 1- and 2-forms at $p$ for which
$\w_{\a}{}^{\c}K_{\c\b}$ is symmetric, then $(\lambda, X, K) \in {\cal
S}_p$).  Given an arbitrary $p\in\M$, each observable function then
determines an element of ${\cal S}_p$.  According to
Thm.~\ref{thm_data}, the converse holds for ordinary quantum
mechanics: there, the algebra of symmetry data at each point $p$ is
{\em integrable}.  In this sense, the phase space of standard quantum
mechanics admits ``as many observables as possible.''  This useful
idea is captured in
\begin{definition}
If, for each $p \in \M$, every element ($\lambda$, $X$, $K$)
$\in {\cal S}_p$ is integrable, we say that ${\cal O}$ is
{\em maximal}.
\end{definition}
\noindent
The set of observables of ordinary quantum mechanics is maximal.  As
we will see, it is essentially this property of the quantum
observables which allows us to recover the standard formalism.

In order to illustrate the importance of the maximality of the space
of observables, it is necessary to introduce one mathematical concept.
It is fairly easy to verify that the Riemann curvature tensor of a
projective Hilbert space is of the special form
\be \label{chsc}
R_{\a\b\c\delta} = \frac{C}{2} \left[
g_{\c[\a} g_{\b]\delta} + \w_{\a\b} \w_{\delta\c} -
\w_{\c[\a} \w_{\b]\delta}  \right],
\ee
where $C = \hbar / 2$ and $[\ldots]$ denotes skew-symmetrization.  If
the curvature tensor of our generalized 
phase space $\M$ satisfies this equation at a point $p$, it is said to
be of {\em constant holomorphic sectional curvature} (CHSC) at $p$.
In that case, the real number $C$ is the value of the holomorphic
sectional curvature at $p$.  If for some real $C$, the Riemann tensor
assumes the form written in \eqn{chsc} (for all $p\in\M$), $\M$ is
called a manifold of CHSC$=C$.  It may be useful to note that if a
K\"ahler manifold $\M$ is of CHSC at each point, then it must be of
overall CHSC; that is, if $R$ satisfies \eqn{chsc} at each $p$, then
$C$ must be a constant \cite{yano}.

Holomorphic sectional curvature, in the context of complex manifolds,
is analogous to the scalar curvature of real manifolds.  Since there
is a very strong relationship between the number of independent
Killing vector fields on a real manifold and the form of the Riemann
curvature tensor \cite{ashtekar}, one may expect that the maximality
of the space of quantum observables to be strongly related to
Eq. (\ref{chsc}).  We will therefore use an approach analogous to that
presented in Ref.~\cite{ashtekar} to consider more closely the
interplay between the geometry of the phase space and the algebra of
observables.  The results in this section are stated without proof;
for details, see Ref.~ \cite{thesis}.

Since the commutator of two Killing vector fields is another Killing
vector field, the set of observables on an arbitrary K\"ahler manifold
is closed under the Poisson bracket $\{,\}$.  The Poisson bracket also
satisfies the Jacobi identity; this equips ${\cal O}$ with the
structure of a Lie algebra.  On the entire function space, we may also
define the commutative operation \be \{f, k\}_+ := (f, k) + fk, \quad
(f, k) := \frac{\hbar}{2} (\grad_\a f) g^{\a\b} (\grad_\b g), \ee
which we call the {\em symmetric bracket}.  The symmetric bracket is
not necessarily closed on the space of observables on an arbitrary
K\"ahler manifold.

Suppose that $f_1$ and $f_2$ are any two observable functions on $\
M$, and let $f_3 = \{f_1, f_2 \}$.  Since each $f_i$ generates a
Killing vector field, it determines an element $(f_i, X_i, K_i)\big|_p
\in {\cal S}_p$ for any $p \in \M$, where $X_i = \hvf{f_i}$ and
$K_i{}_{\a\b} = \grad_\a(\hvf{f_i})_\b$.  Of course, $f_3 = \w(X_1,
X_2)$ and $X_3{}_\a = -g_{\a\b} \big[\hvf{f_1}, \hvf{f_2}\big]^\b
\big|_p = X_2^\b K_1{}_{\b\a} - X_1^\b K_2{}_{\b\a}$, where $[,]$
denotes the commutator of vector fields.  It is straight-forward to
derive the corresponding expression for the 2-form $K_3$; the result
is \be K_3{}_{\a\b} = K_2{}_\a{}^\c K_1{}_{\c\b} - K_1{}_\a{}^\c
K_2{}_{\c\b} + X_1{}_\mu X_2{}_\nu R_{\a\b}{}^{\mu\nu}.  \ee This fact
suggests that we define the following operation on the set ${\cal
S}_p$ of symmetry data: 
\be \label{pb_p}
\begin{array}{ll}
	\left[ (f_1, X_1, K_1), (f_2, X_2, K_2) \right]_p :=
	\bigg(&
		\w(X_1, X_2), \\
		&  X_2^\b K_1{}_{\b\a} - X_1^\b K_2{}_{\b\a}, \\
		&  K_2{}_\a{}^\c  K_1{}_{\c\b} -
		   K_1{}_\a{}^\c  K_2{}_{\c\b}
		   + X_1{}_\mu X_2{}_\nu R_{\a\b}{}^{\mu\nu} \bigg).
\end{array}
\ee
This bracket on ${\cal S}_p$ is defined only to mirror the Poisson
bracket on ${\cal O}$.  That is, to obtain the symmetry data
corresponding to the Poisson bracket of two observables, one may
simply apply the above ``bracket'' operation to the symmetry data of
the initial observables. 

For an arbitrary point $p$ on an arbitrary K\"ahler manifold, ${\cal
S}_p$ is closed under $[\, ,\,]_p$.  However, the Jacobi identity
will, in general, fail.  It is natural to ask for circumstances under
which ${\cal S}_p$ forms a Lie algebra.  The answer to this question
is provided by
\begin{lemma} \label{lem_jacobi} 
$[\, ,\, ]_p$ is a Lie bracket on ${\cal S}_p$ if and only if the
Riemann tensor is of CHSC at p.
\end{lemma}
\noindent
If ${\cal O}$ is maximal, then $[ , ]_p$ is a Lie bracket on ${\cal
S}_p$ for any $p\in\M$.  As a consequence of Lemma~\ref{lem_jacobi},
the Riemann tensor is of CHSC at each point of $\M$.  Therefore $\M$
is a manifold of CHSC; this is summarized by
\begin{cor}\label{cor2}
Suppose ${\cal O}$ is maximal.  Then $\M$ is a manifold of CHSC.
\end{cor}

Let us make the analogous construction for the symmetric bracket.
Again, let $f_1, f_2 \in {\cal O}$ and let $f_4$ denote the symmetric
bracket, $f_4 = \{f_1, f_2 \}_+ = f_1 f_2 + (\hbar/2) g(X_1, X_2)$. 
One can easily show the following:
\ba
	\w_\a{}^\b \grad_\b f_4 &=&
	f_1 X_2{}_\a + f_2 X_1{}_\a + 
	(\hbar/2) \w_\a{}^\b\left(
	K_1{}_{\b\c} X_2^\c + K_2{}_{\b\c} X_1^\c \right),\\
	\grad_\a X_4{}_\b &=& 
	f_1 K_2{}_{\a\b} + f_2 K_1{}_{\a\b} +
	\hbar K_1{}_{\c[\a} \w^{\c\delta} K_2{}_{\b]\delta} +
	X_1^\mu X_2{}_\nu\left[
	\hbar R_{\a(\mu\nu)\c} - 2 g_{\a(\mu} g_{\nu)\c}
	\right] \w_\b{}^\c.
\ea
Therefore it is natural to define the following commutative operation
on ${\cal S}_p$:
\be\label{sb_p}
\begin{array}{ll}
	\big((f_1, X_1, K_1), (f_2, X_2, K_2) \big)_p :=
\bigg( &

	f_1 f_2 + \hbar/2 g( X_1, X_2 ), \\
&	f_1 X_2{}_\a + f_2 X_1{}_\a + 
	(\hbar/2) \w_\a{}^\b\left(
	K_1{}_{\b\c} X_2^\c + K_2{}_{\b\c} X_1^\c \right),\\
&	f_1 K_2{}_{\a\b} + f_2 K_1{}_{\a\b} +
	\hbar K_1{}_{\c[\a} \w^{\c\delta} K_2{}_{\b]\delta}+ \\
&       X_1^\mu X_2{}_\nu  \left[
	\hbar R_{\a(\mu\nu)\c} - 2 g_{\a(\mu} g_{\nu)\c}
		\right] \w_\b{}^\c  \bigg).
\end{array}
\ee
We know that if ${\cal M}$ is a projective Hilbert space this
operation produces the symmetry data determined by the symmetric
(i.e. Jordan) bracket of the corresponding observables.  For the
generic case however, ${\cal O}$ will not be closed under the
symmetric bracket; hence we do not expect ${\cal S}_p$ to be closed
under its symmetric bracket.  Therefore, the symmetric bracket defined
by \eqn{sb_p} should be viewed as an operation on the space of {\em
all} triples $(f, X, K)$, without the symmetry condition imposed on
$K$.

The condition that ${\cal S}_p$ be closed under $( , )_p$ is even stronger
than that found in Lemma~\ref{lem_jacobi}:
\begin{lemma}\label{lem_sb}
The set ${\cal S}_p$ of symmetry data at $p$ is closed under
the symmetric bracket $( , )_p$ if and only if the Riemann
tensor is of CHSC$=2/\hbar$ at p.
\end{lemma}
The difference between this condition and that specified in
Lemma~\ref{lem_jacobi} is that here the actual value of the
holomorphic sectional curvature is determined by the coefficient
appearing in the definition of the symmetric bracket.
Lemma~\ref{lem_jacobi} states that the holomorphic sectional curvature
be constant, but does not specify its value.

Note that each of these lemmas involves only a single point of $\M$.
If the set of observables is maximal, then by definition its elements
are in one-to-one correspondence with elements of ${\cal S}_p$ for any
$p\in\M$.  Therefore, the closure of a maximal set of observables
under the symmetric bracket is equivalent to the closure of ${\cal
S}_p$ under $( , )_p$ for any $p\in\M$.  If we apply
Lemma~\ref{lem_sb} to each point of $\M$, we immediately obtain
\begin{cor}
Suppose ${\cal O}$ is maximal.  Then ${\cal O}$ is closed under
the symmetric bracket if and only if the Riemann tensor is
of CHSC$=2/\hbar$.
\end{cor}

Corollary \ref{cor2}  does not specify the value of the holomorphic
sectional curvature, but merely states that the Riemann tensor assumes
the special form written in \eqn{chsc} for some constant $C$.  However,
combining it with Lemma~\ref{lem_sb}, one obtains
\begin{theorem}
Suppose ${\cal O}$ is maximal and that ${\cal S}_p$ is
closed under $( , )_p$ for a single point $p\in\M$.  Then
$\M$ is a manifold of CHSC$=2/\hbar$.

\end{theorem}

Of course, we would like to go one step further to say that $\M$ must
be a projective Hilbert space. Now, it is known that any two
finite-dimensional K\"ahler manifolds which are complete,
simply-connected and of CHSC$=2/\hbar$ are isomorphic \cite{yano}.
Therefore, in the finite-dimensional case, we have obtained a
characterization of the structure that picks out the standard quantum
kinematics from possible generalized frameworks: If the generalized
quantum phase space is a complete, simply-connected K\"ahler manifold
and the set of observables is maximal and closed under $\{ , \}_+$,
then one is dealing with the structure of ordinary quantum
mechanics. This characterization should be useful in the search for
genuine generalizations. In a generalized theory, one or more of the
conditions must be violated. Thus, the characterization systematizes
the search for generalizations and suggests concrete directions to
proceed.

The case when $\M$ is infinite-dimensional is much more interesting
physically. However, in this case, we do not know if we can conclude
that a (complete) K\"ahler manifold on which ${\cal O}$ is maximal and
closed under $\{ , \}_+$ is isomorphic to a projective Hilbert space.
Indeed there may exist {\em many} different infinite-dimensional
K\"ahler manifolds (satisfying the above completeness requirements) of
the same constant holomorphic sectional curvature. This is an
important open problem. If the situation turns out to be the same as
in the finite-dimensional case, we will again have a theorem
characterizing ordinary quantum mechanics. If the situation is
different and we have many such K\"ahler manifolds, it would be even
more interesting. For, we would then be presented with viable
generalizations of the standard quantum formalism. Such examples would
be very interesting because they are likely to admit a consistent
measurement theory and thus lead to physically complete
generalizations of a rather subtle type.

\section{Semi-classical considerations} \label{sec4}

One of the most striking features of the geometric approach to quantum
mechanics is its resemblance to the classical formalism. One might
therefore expect that the geometric framework may shed some light on
the relation between quantum and classical physics and, in particular,
semi-classical approximations. We will see that this expectation is
correct. In section \ref{sec4.A} we consider the relation between the
two theories at a kinematical level and elucidate the special role
played by coherent states. These results are then used in the
remaining section to analyze dynamical issues. In \ref{sec4.B} we
consider systems such as harmonic oscillators and free fields; in
\ref{sec4.C}, we discuss dynamics in the WKB approximation.

\subsection{Kinematics} \label{sec4.A}

Let us now suppose that we are given a classical phase space $(\Gamma,
\a)$, where $\Gamma$ is assumed to be a finite-dimensional vector
space and $\a$ is the symplectic form thereon. Let $(\P, g, \w)$ be
the quantum phase space which results from the application of the
text-book quantization procedure to $(\Gamma, \a)$. Thus, on the
quantum Hilbert space $\H$, there are position and momentum operators
($\hat{Q}_i$ and $\hat{P}_i$, $i=1, ..., n$) corresponding to the
classical position and momentum observables.  We are all accustomed to
this direction of construction. Our goal now is to do things in
``reverse'': Given the quantum phase space, we will provide a
construction of the classical one.

We will continue with our previous convention and write the
projections to $\P$ of the expectation values of these operators as
$q_i$ and $p_i$. It should be emphasized, however, that these
functions are the elementary {\em quantum} observables, not the
classical ones; they are functions on the infinite-dimensional space
$\P$ and not on the finite-dimensional space $\Gamma$. Note that by
standard construction the Poisson brackets of these elementary quantum
observables satisfy the same ``commutation relations'' as do the
classical variables:
\be \{q_i,
p_j\} = \delta_{ij} \quad {\rm and} \quad \{q_i, q_j\} = 0 = \{p_i,
p_j\}.  
\ee

One of the most common approaches to the classical limit comes from a
simple theorem of Ehrenfest's which suggests that expectation values of
quantum operators are to be approximated, in some sense, by the
corresponding classical observables.  The geometric formulation is
particularly well-suited to implement these ideas.  To each
quantum state $x\in\P$, let us associate the classical state $(q_i(x),
p_j(x)) \in \Gamma$.  This association defines the obvious mapping
$\rho:\P \rightarrow \Gamma$.  In fact, one might view this as the
{\em definition} of the classical phase space.  (The astute reader will
object, correctly pointing out that one must specify the elementary
quantum observables $q_i$ and $p_i$ before such a ``definition'' of
the classical phase space can be made. Recall, however, that this is
just the counterpart of the fact that elementary classical observables
must be specified before construction of the quantum theory.)

For convenience, let us denote the generic elementary observables by
$f_r$, $r=1 \ldots 2n$ (i.e., $\{f_r\} = \{q_i, p_i, i=1 \ldots n\}$).
Now let us define an equivalence relation on $\P$: $x_1 \sim x_2 \;
\iff \; f_r(x_1)=f_r(x_2) \; \forall r$.  This equivalence relation
fibrates the quantum phase space, which may now view as a bundle over the
classical phase space, $\Gamma = \P/\sim$.

Associated to any bundle is a ``vertical'' distribution; i.e., a
special class of (vertical) tangent vectors at each state $x\in\P$.
That a tangent vector $v \in T_x\P$ is vertical simply means that
$v(f_r) = 0 \; \forall r=1 \ldots 2n$.  Equivalently the vertical
subspace may be defined as
\be
{\cal V}_x := \big\{ v\in T_x \P \; \big| \;
\w(\hvf{f_r}|_x, \; v) = 0 \; \forall r=1 \ldots 2n \big\} .
\ee
The vertical directions are simply those in which the elementary
quantum observables assume constant values.

Let ${\cal V}_x^\perp$ denote the $\w$-orthogonal complement of the
vertical subspace at $x$.  One can show \cite{thesis} that each
tangent space $T_x\P$ may be written as the sum $T_x\P = {\cal V}_x
\oplus {\cal V}_x^\perp$.  Let us therefore call elements of ${\cal
V}_x^\perp$ {\em horizontal}.  If $Y$ is a vector {\em field} which is
everywhere horizontal (resp. vertical), we will simply write $Y \in
{\cal V}^\perp$ (resp. $Y \in {\cal V}$).  Note in particular that any
algebraic combination of the $\hvf{f_r}$ is necessarily horizontal
everywhere.  Let us make three preliminary observations that will be
used later.  First, the distribution ${\cal V}$ is {\em integrable}
since for any $v_1, v_2 \in {\cal V}, \; [v_1, v_2](f_r)=0 \implies
[v_1, v_2] \in {\cal V}$.  Next, since $\hvf{f_r}(f_s) = \{f_s,
f_r\}$, we must have $[\hvf{f_r}, v](f_s) = 0 \; \forall v\in{\cal
V}$.  Therefore, the $\hvf{f_r}$ preserve the vertical distribution.
Similarly, the Hamiltonian vector field of any algebraic function of
the $f_r$ also preserves the distribution.  Lastly, since $[\hvf{f_r},
\hvf{f_s}] = 0 \; \forall r,s$ the horizontal spaces are also
integrable.  Therefore, there exist global horizontal sections of our
quantum bundle over $\Gamma$.

We are now prepared to reconstruct the classical symplectic structure
from the geometry of the quantum phase space.  Let $\xi$ and $\zeta$
be two vector fields on $\Gamma$ and denote by $\tilde{\xi}$ and
$\tilde{\zeta}$ their horizontal lifts to $\P$.  Then the classical
symplectic structure is defined by \be \label{omega_defined} \a(\xi,
\zeta) := \w(\tilde{\xi}, \tilde{\zeta}).  \ee We have simply defined
$\a$ as the ``horizontal part'' of the quantum symplectic structure.
Since the classical phase space is linear, it is obvious that
\eqn{omega_defined} correctly defines the classical symplectic
structure.  However, in order to allow more general constructions, let
us see why this definition of $\a$ provides a symplectic structure on
$\Gamma$.  First of all, it is easy to show that $\w(\tilde{\xi},
\tilde{\zeta})$ is constant along the fibres of $\P$.  Therefore, the
definition is self-consistent.  To see that $\a$ is non-degenerate,
one need only notice that $\a(\xi, \zeta)=0 \; \forall \zeta \implies
\w(\tilde{\xi}, \tilde{\zeta}) = 0 \; \forall \tilde{\zeta} \in {\cal
V}^\perp \implies \tilde{\zeta} \in {\cal V}$; by construction,
$\zeta$ must therefore vanish.  Finally, that $\a$ is closed is
obvious since it is the pull-back, via any horizontal section, of a
closed form.

Let us summarize these results.  Quantization of a classical theory
with a {\em linear} phase space is canonical, given the elementary
position and momentum observables.  For this case, the quantum phase
space may be naturally viewed as a bundle over the classical phase
space.  The classical phase space is simply the base space of this
bundle.  The symplectic structure on $\P$ naturally defines a notion
of horizontality, and the classical symplectic structure is simply the
horizontal part of $\w$.

Incidentally, note that our arguments go a short step toward the more
general theory for which the classical phase space is not necessarily
linear.  Suppose that ${\cal V}$ is an integrable, symplectic
distribution on $\P$ of finite co-dimension $2n$ which may be
specified {\em locally} by the constancy of functions $f_r, r=1 \ldots
2n$.  Suppose also that these functions may be chosen in such a way
that their Poisson algebra closes (up to constants).  Then the
quotient of $\P$ by ${\cal V}$ inherits a natural symplectic
structure.  It is desirable to eliminate the requirement that the
co-dimension of ${\cal V}$ be finite; otherwise treatment does not
apply to, e.g., quantum field theory.  We believe that this assumption
can be removed although a detailed analysis is yet to be carried out.

Note that our construction provides not only a projection from $\P$ to
$\Gamma$ but also a class of preferred embeddings of classical phase
space into the quantum one.  One may guess that these embeddings are
related, in some way, to coherent states.  After all, it is well-known
that spaces of coherent states are endowed with natural symplectic
structures \cite{zhang,perelomov}.

Let us therefore examine the nature of the coherent state spaces from
the geometric point of view.  There are a number of different
constructions of coherent states.  We consider the attractive and
fairly general approach introduced by Perelomov\cite{perelomov1},
Gilmore\cite{gilmore} and, to some extent much earlier, by
Klauder\cite{klauder}. Consider the the Heisenberg-Weyl group,%
\footnote{Although this group is kinematical in origin, in
the literature on coherent states, it is often referred to as the
{\em dynamical group} associated with the simple harmonic oscillator.}
obtained by exponentiating the Lie algebra generated by $\hat{Q_i}$,
$\hat{P_j}$ and the identity operator on $\H$.  One defines
Perelomov's space of generalized coherent states by the action of this
group on an arbitrary element $\Psi_0 \in \H$. In this manner, one
obtains generalized coherent states $\Psi_{(q^\prime_i, p^\prime_i)}
:= \exp[ -\frac{i}{\hbar}\sum_i(q'_i\hat{P}_i - p'_i\hat{Q}_i)]\, \Psi_0$,
labeled by pairs of parameters $(q^\prime_i, p^\prime_i)$ with
dimensions of position and momenta, respectively.  If one chooses for
$\Psi_0$ the ground state of the oscillator, one recovers the space of
standard coherent states.  However, one may apply the above
construction using arbitrary $\Psi_0$ as the fiducial state. Hence,
the notion of generalized coherent states is a viable one for the
generic system.

Let us now return to the relation between the quantum and classical
phase spaces. Given an arbitrary element $x_0 \in \P$ of the quantum
phase space, one can obtain a sub-manifold of generalized coherent
states as follows: Choose a state $\Psi_0$ which projects to $x_0$ and
construct the generalized coherent states via the standard method
described above. To begin with, these states constitute a genuinely
non-linear subspace of the Hilbert space. Nonetheless, we may project
the entire space to $\P$. An understanding of the nature of the
resulting sub-manifold of $\P$ is provided by the following
facts. First, the expectation values of the basic operators at the
coherent states are given by
\be
  q_i( x_{(q^\prime, p^\prime)} ) = q_i( x_0 ) + q^\prime_i
	\quad {\rm and} \quad
  p_i( x_{(q^\prime, p^\prime)} ) = p_i( x_0 ) + p^\prime_i,
\ee
(so that if we choose for $\Psi_0$ the ground state of the harmonic
oscillator, the expectation values are precisely $q'$ and $p'$.)
Second, any two coherent states generated by the same fiducial element
$x_0$ possesses the same uncertainties. Therefore, the uncertainties
in $q_i$ and $p_j$ are constant on the generalized coherent state
space, and hence on the sub-manifold of $\P$ they define. Finally, 
note that the definition of the Heisenberg-Weyl group
implies:
\ba \frac{\partial}{\partial q^\prime}
\Psi_{(q^\prime, p^\prime)} = \frac{1}{\i\hbar} \hat{P}
\Psi_{(q^\prime, p^\prime)}, \\ \frac{\partial}{\partial p^\prime}
\Psi_{(q^\prime, p^\prime)} = - \frac{1}{\i\hbar} \hat{Q}
\Psi_{(q^\prime, p^\prime)}.  
\ea 
Projecting this result to $\P$, one immediately verifies that each
space of generalized coherent states is everywhere horizontal. Thus,
our horizontal sections on $\P$ are precisely the generalized coherent
state spaces! Hence, it follows in particular that the uncertainties
in the elementary quantum observables are constant on the horizontal
sections.

Finally, it is interesting to note that our construction of the
horizontal sections did {\em not} refer to the Heisenberg group.
Given a general system, Perelomov's generalized coherent state spaces
are constructed as above, but by use of a different (``dynamical'',
see footnote 8) group; an entirely different set of
generalized coherent state spaces will then result, which will {\em
not} correspond to horizontal sections.  This seems an a point worth
further investigation.

\subsection{Dynamics: Oscillators} \label{sec4.B}

Let us now consider dynamical issues.  Since the horizontal spaces are
integrable, the classical phase space may be embedded into $\P$ in
infinitely many ways.  Is there a {\em preferred} horizontal section?
The answer to this question will involve the dynamics of the system.
In general, we expect this issue to be far from trivial.  The case of
the harmonic oscillator is, as one may guess, fairly simple. In this
sub-section we will restrict ourselves to this case. 

Recall the elementary treatment of the classical limit in terms of
Ehrenfest's theorem.  The rate of change of the expectation value
$\< \hat{F} \>$ of any observable operator $\hat{F}$ is given by
\be \label{ehrenfest}
\frac{\d}{\d t} \< \hat{F} \> = \frac{1}{\i\hbar} 
\left<\left[ \hat{F}, \hat{H} \right] \right>,
\ee
where $\hat{H}$ is the Hamiltonian operator.  As usual, let $f$ and
$h$ denote the corresponding observable functions on $\P$.  Then as a
result of \eqn{pb_on_P} and its subsequent discussion, \eqn{ehrenfest}
directly translates to
\be
	\frac{\d f}{\d t} = \{ f, h \}.
\ee

The fact that this equation exactly mirrors the classical expression
is, however, deceiving because the functional form of the quantum
Hamiltonian $h$ is typically entirely different from that of the
classical Hamiltonian.  For example, in general, one may not even be
able to express $h$ in terms of $q_i$ and $p_i$ alone. In fact, this
is the case already for the harmonic oscillator. At first, this seems
puzzling because, as is well-known, the $q_i$ and $p_i$ ``follow the
classical trajectories'' if the Hamiltonian operator is a quadratic
function of $\hat{Q_i}$ and $\hat{P_i}$.

Let us therefore pursue this a bit further. For simplicity,
consider the one-dimensional case.  First, we must obtain the form of
the quantum Hamiltonian.  This is easy.  Since $ \hat{H} = (1 /
2m)\hat{P}^2 + (m\w^2 / 2)\hat{Q}^2 = (1 / 2m) \{ \hat{P}, \hat{P}
\}_+ + (m\w^2 / 2) \{ \hat{Q}, \hat{Q} \}_+$, the Hamiltonian function
on $\P$ must be of the form
\be h = \frac{1}{2m}p^2 + \frac{m\w^2}{2}q^2 +
\frac{1}{2m}(\Delta p)^2 + \frac{m\w^2}{2}(\Delta q)^2, \ee 
where we have used only the definition and properties of the symmetric
bracket.  Cross-terms involving the uncertainties---and hence the
total Hamiltonian $h$---can not be expressed in terms of $q_i$ and
$p_i$ alone. More generally, such cross terms are by-products of the
quantization process and provide one of the significant differences
between the forms of the classical and quantum dynamics.

Notice that the Hamiltonian may be decomposed into two parts; hence,
so may the corresponding Hamiltonian vector field.  $\hvf{h} =
\hvf{h_0} + \hvf{h_\Delta}$ where $h_0$ takes the form of the
classical Hamiltonian and $h_\Delta = (1/2m)(\Delta p)^2 +
(m\w^2/2)(\Delta q)^2$.  Now, recall that the uncertainties $\Delta q$
and $\Delta p$ are constant in the horizontal directions.  As a
consequence, $\hvf{h_\Delta}$ must be purely vertical.  This should
not be at all surprising since, as we already know, time-dependence of
$q$ and $p$ is the same as for the classical observables.  Since
$\hvf{h_\Delta}$ is everywhere vertical, the pull-back $s^* h$ of the
quantum Hamiltonian, via any horizontal section $s : \Gamma
\rightarrow \P$, to the classical phase space is precisely the
classical Hamiltonian, up to a physically irrelevant overall constant.
This is essentially the reason why the evolution of the basic quantum
observables agrees with the classical evolution.  However, the quantum
evolution is quite different from the classical, for it does not
generally preserve the horizontal sections.  A natural question
arises: is there a horizontal section which is preserved by the
Hamiltonian evolution?  

Note that we are asking for much more than a dynamical trajectory
confined to one horizontal section. The question is whether there
exists an {\it entire} horizontal section of $\P$ which is preserved
by the quantum evolution.  A section $s_0 : \Gamma \rightarrow \P$ is
of this special nature if and only if $\hvf{h_\Delta}$ vanishes on the
entire image of $s_0$.  Equivalently, we may search for a section
$s_0$ on which the uncertainty term $h_\Delta$ attains a local
extremum.  It seems a rather natural guess that this will be the case
at states which saturate the uncertainty relation between $q$ and $p$.
This expectation is correct.  One can see this by writing
\be 
(1/\w\hbar) h_\Delta \, = \, \left[ \Delta q \sqrt{m\w/2\hbar} -
\Delta p \sqrt{1/2m\w\hbar} \right]^2 + (1/\hbar) \Delta p \Delta q
\,\, \ge 1/2, 
\ee 
where the inequality is due to Heisenberg.  Therefore, any state $x$
at which $h_\Delta(x) = \w\hbar/2$ extremizes the uncertainty term.
It is now easy to see that the only values of $\Delta p$ and $\Delta
q$ (subject, of course, to Heisenberg's uncertainty relation) which
extremize $h_\Delta$ are given by $(\Delta q)^2 = \hbar/2m\w$ and
$(\Delta p)^2 = m\w\hbar/2$.  There is therefore a {\it single}
horizontal section which is preserved by the Hamiltonian evolution; it
is the section on which the quantum evolution is ``most classical''.
As one might suspect, this preferred section corresponds to the
standard coherent state space, generated by the oscillator's ground
state.  These results clearly hold for any finite-dimensional
oscillator, and likely for the infinite-dimensional case (e.g.,
quantum Maxwell theory).

\subsection{Dynamics: WKB approximation} \label{sec4.C}

The previous discussion of dynamics has been, to some extent, in the
context of the Ehrenfest approach to semi-classical dynamics.  Let us
also consider the problem from the point of view of Hamilton-Jacobi
theory---the context in which one often introduces the WKB
approximation.  While the discussion of section \ref{sec4.B} has been
limited to the special case of the oscillator, we now consider more
general systems and obtain two main results.  First, we will present
interesting condition for the validity of the WKB approximation which,
to our knowledge, has not been discussed in the literature.  Second,
we will show that the WKB equation actually corresponds to a
Hamiltonian evolution on the projective space, and therefore defines a
generalized quantum dynamics of the Weinberg type.

We will consider the dynamics of a non-relativistic particle moving in
$\R^n$ under the influence of a general conservative force (see
footnote 6). The wave function is therefore assumed to satisfy the
Schr\"odinger equation
\be
\i\hbar \frac{\partial}{\partial t} \Psi(x,t) =
\left( -\frac{\hbar^2}{2m}\Delta + V(x) \right) \Psi(x,t),
\ee
where $\Delta$ is the Laplace operator on $\R^n$.  In the spirit of
our approach, we decompose the state vector into real and imaginary
parts.  This defines two real fields $\phi$ and $\pi$ via $\Psi =
(\phi + \i\pi)/\sqrt{2\hbar}$.  In terms of these fields, the metric
and symplectic structure on $\H$ assume the forms
\ba
G( (\phi_1, \pi_1), (\phi_2, \pi_2) ) =  \int \d^n \! x 
\left[ \phi_1(x) \phi_2(x) + \pi_1(x) \pi_2(x) \right], \\ 
 \W( (\phi_1, \pi_1), (\phi_2, \pi_2) ) = \int \d^n \! x 
\left[ \phi_1(x) \pi_2(x) - \phi_2(x) \pi_1(x) \right].
\ea
The second equation may look familiar from classical field theory.  It
implies that the fields $\phi$ and $\pi$ are canonically conjugate;
one may view them, respectively, as the `field' and `momentum density'
of the classical field theory under consideration.  This is not
surprising because, after all, any quantum theory can be regarded as a
field theory.

Indeed, one may calculate the expectation value of the above
Hamiltonian operator,
\be
H(\phi, \pi) = \frac{1}{2\hbar} \int \d^n \! x  \left[
\frac{\hbar^2}{2m} \left( (\vec{\partial}\phi)^2 
+ (\vec{\partial}\pi)^2 \right)
+ V(x)\left( \phi^2(x) + \pi^2(x) \right) \right],
\ee
and verify that the fields evolve according to the canonical equations
of motion:
\be \label{eqs-of-motion}
\frac{\partial \phi}{\partial t} = \frac{\delta H}{\delta \pi}
\quad {\rm and} \quad
 \frac{\partial \pi}{\partial t} = - \frac{\delta H}{\delta \phi}.
\ee
We are interested in the relationship between these equations of
motion provided by quantum dynamics and the Hamilton-Jacobi equation.

For this it is convenient to express $\Psi$ as $\Psi =
\sqrt{\rho}\exp( \i S/\hbar )$ and rewrite Schr\"odinger dynamics as
equations of motion for $\rho$ and $S$:
\ba\label{ham-jac}
\frac{\partial S}{\partial t} + \frac{1}{2m}(\vec{\partial}S)^2
+ V(x)&=& \frac{\hbar^2}{2m} \frac{\Delta \sqrt{\rho}}
{\sqrt{\rho}},\\
m\frac{\partial \rho}{\partial t} + \vec{\partial} \cdot
(\rho \vec{\partial} S) &=& 0.
\ea
The second equation is the conservation equation, $\partial \rho /
\partial t + \vec{\partial}\cdot \vec{J} = 0$, where $\vec{J} =
\rho\vec{\partial}S/m$ is the probability current density.  The
familiar observation is the fact that \eqn{ham-jac} becomes the
Hamilton-Jacobi equation upon dropping the term involving $\hbar$; in
a loose sense, one obtains the classical limit upon taking $\hbar
\rightarrow 0$. (For details, see standard texts, e.g.,
\cite{goldstein,lanczos}.)

Let us explore the standard approach in which one simply drops the
``quantum correction'' to the Hamilton-Jacobi equation.  To be
concrete, let us refer to the corresponding evolution
as {\it WKB evolution}.  First, one may
express the Hamiltonian function in terms of $\rho$ and $S$:
\be \label{hamiltonian}
H(\rho, S) = \int \d^n\!x \left[ \frac{\hbar^2}{8m\rho(x)}
(\vec{\partial}\rho)^2
+ \frac{1}{2m}\rho(x) (\vec{\partial} S)^2 + \rho(x)V(x)\right].
\ee
It may be seen that $\rho$ and $S$ are also canonically related and
that their equations of motion may be obtained from
Eqs. (\ref{eqs-of-motion}) by making the replacements $\phi
\rightarrow \rho$ and $\pi \rightarrow S$.  One may also verify that
the WKB evolution corresponds to dropping the first term in
Eq. (\ref{hamiltonian}).  That is, if $H_\hbar$ denotes this first
term, and $H_{WKB} := H - H_\hbar$, then the WKB evolution corresponds
to integration of the Hamiltonian vector field $X_{WKB}$ generated by
$H_{WKB}$.

Let us first examine the condition of validity of the the WKB
approximation.  We require conditions under which the
Hamiltonian vector field generated by $H_\hbar$ is small compared to
that generated by $H$.  To this end, we compute the functional
derivatives of $H_\hbar$ with respect to the fields $\phi$ and $\pi$.
We obtain
\be \label{funct_deriv}
\frac{\delta H_\hbar}{\delta \phi} = 
-\frac{1}{8m\hbar\rho^2} \left[2\hbar\rho\left( \phi\Delta\phi 
+\pi\Delta\pi \right) + \left( \phi \vec{\partial}\pi -
\pi\vec{\partial}\phi \right)^2\right] \phi;
\ee
the corresponding expression for $\delta H / \delta\pi$ is obtained by
making the replacement $\phi \leftrightarrow \pi$ above.  This
expression may be written in terms of quantities which are somewhat
more physical.  One need only notice that
\be
\begin{array}{rlrl}
\left< \hat{P} \right> &= \int \d^n \! x \; m\vec{J}(x), \quad
\quad& \vec{J} &= \frac{1}{2} ( \phi\vec{\partial}\pi -
\pi\vec{\partial}\phi ), \\
\left< \hat{P}^2 \right> &= \int \d^n \! x \; K(x), \quad
\quad& K &= \frac{\hbar}{2} ( \phi\Delta\phi +
\pi\Delta\pi ).
\end{array}
\ee
The quantities $m\vec{J}$ and $K$ which appear in
Eq. (\ref{funct_deriv}) may be interpreted physically as the ``density
of momentum'' and the ``density of squared-momentum''.

In terms of these quantities, the above functional derivative may be
written as
\be
\frac{\delta H}{\delta \phi} = -\frac{\1}{2m\hbar\rho^2} \left[
(m\vec{J})^2 - \rho K \right] \phi.
\ee
Therefore,
\be
\i\hbar\left( \frac{\delta H_\hbar}{\delta \pi} - \i
\frac{\delta H_\hbar}{\delta \phi} \right) =
\frac{\i}{2m\rho^2}\left( (m\vec{J})^2 - \rho K \right)
(\phi + \i\pi). 
\ee
The WKB evolution is obtained by subtracting ($(1/\sqrt{2\hbar}$
times) this term from the Schr\"odinger equation.  The WKB equation of
motion may therefore be written
\be\label{wkb}
\i\hbar \frac{\partial}{\partial t}\Psi = \hat{H}\Psi + 
\frac{1}{2m\rho^2}\left[(m\vec{J})^2 - \rho K \right] \Psi,
\ee
where $\hat{H}$ is the unaltered Hamiltonian operator. Hence, the
condition of validity of the WKB equation is that the second term on
the right side of Eq. (\ref{wkb}) be small; i.e. that the ``density of
squared-momentum'' (weighted by the probability density) be comparable
in magnitude to the square of the ``density of momentum''.

We have just arrived at an interesting form of the WKB equation of
motion. One should compare it with the general form given by
Eq.~(\ref{gen_nonlin}) of the non-linear Schr\"odinger equation. Note
that the additional term $\a$ appearing in that equation is a function
of only the amplitude $\rho$. Hence, the WKB equation appears to be
considerably more complicated than the explicit forms of generalized
dynamics that are generally considered.

However, we have already seen (c.f. the discussion following
Eq. (\ref{hamiltonian})) that the WKB evolution is generated by a
Hamiltonian function on $\H$. Surprisingly, this evolution actually
induces a Hamiltonian flow on the projective Hilbert space.  To see
this, in the light of section \ref{sec3.A}, we need only show that
$X_{WKB}$ preserves the unit sphere.  This is the case if and only if
the Poisson bracket of $H_{WKB}$ and the constraint function, $C =
\int \d^n \!x \rho(x) -1$ vanishes at each normalized state. This fact
may be proven as follows: Since $H_{\hbar}$ is independent of $S$,
$\{H_\hbar, C \} = 0$.  Moreover, since $H$ generates the
Schr\"odinger evolution, which preserves the unit sphere, $\{H,
C\}=0$.  Taking the difference between these two Poisson brackets, we
obtain the desired result: $\{H_{WKB}, C\} = 0$.  This Poisson bracket
actually vanishes {\em strongly}; therefore, the WKB evolution
actually preserves the norm of {\em all} state-vectors (not just the
unit vectors).  Therefore, the dynamics of the WKB approximation is an
example of generalized dynamics of the Weinberg type; in particular,
it defines a Hamiltonian flow on the projective Hilbert space.

\section{Discussion}

Let us begin with a summary of the main results. 

We first showed that ordinary quantum mechanics can be reformulated in
a geometric language.  In particular, as in classical mechanics, the
space of physical states---the ``quantum phase space''---is a
symplectic manifold and dynamics is generated by a Hamiltonian vector
field. However, unlike in classical mechanics, the quantum phase space
is equipped also with a K\"ahler structure. As one might suspect, the
Riemannian metric, which is absent in the classical description,
governs uncertainty relations and state vector reduction which are
hallmarks of quantum mechanics. The geometric formulation shows that
the linear structure which is at the forefront in text-book treatments
of quantum mechanics is, primarily, only a technical convenience and
the essential ingredients---the manifold of states, the symplectic
structure and the Riemannian metric---do not share this
linearity. Therefore, the framework can serve as a stepping stone
for non-linear generalizations of quantum mechanics.

One can consider such generalizations in various directions. A
``conservative'' approach would retain the kinematical structure and
generalize only the dynamics. This is the viewpoint that underlies the
various non-linear generalizations of the Schr\"odinger equation that
have been considered in the literature. The strategy is also natural
and easy to implement in the geometric formulation: While in the
standard formulation, dynamics can be generated by a very restricted
class of functions on the quantum phase space, we can generalize the
framework by allowing the Hamiltonian flow to be generated by {\it
any} smooth function on the quantum phase space. We saw that the known
proposals of generalized dynamics fall in this class. Furthermore, by
concentrating on the quantum phase space---rather than the fiducial
Hilbert space---one can separate essential features of dynamics from
inessential ones. In particular, this naturally led to a clarification
of the relation between Weinberg's framework and non-linear
Schr\"odinger equations and corrected some misconceptions.

The kinematical generalizations are more difficult. However, we were
able to streamline the search for such generalizations by providing a
characterization of ordinary quantum mechanics using structures that
have direct physical interpretation: In essence, what singles out
quantum mechanics is that the underlying K\"ahler manifold has maximal
symmetries. {}From the geometric viewpoint then, restricting oneself
to ordinary quantum mechanics is rather analogous, in
(pseudo-)Riemannian geometry, to working {\it only} with manifolds of
constant curvature. Now, in (pseudo-)Riemannian geometry, a rich
theory remains even if one drops the restriction of maximal
symmetries. Indeed, in the context of general relativity, the
manifolds of constant curvature are only the `vacua' and in most
physically interesting situations, there are significant departures
from these geometries. Is the situation perhaps similar in the case of
quantum mechanics? Have we been restricting our attention only to the
most elementary of viable theories?

The geometrical formulation is also useful to probe semi-classical
issues. We saw in particular, that the quantum phase space has a
natural bundle structure and the horizontal cross-sections correspond
precisely to families of generalized coherent states.  It also
provides a succinct and clear condition for validity of the WKB
approximation. Furthermore, it turned out that dynamics in the WKB
approximation yields a well-defined flow on the quantum phase space
which corresponds to a ``generalized quantum dynamics'' in the sense
of Weinberg.

The quantum (Hilbert space and hence) phase space is
finite-dimensional only in exceptional cases, such as spin
systems. Most work in the literature in the area of geometric
formulations of quantum mechanics deals only with this case.  By
working with Hilbert manifolds, we were able to treat the generic
case---such as particles moving in $\R^n$---where the quantum phase
space is infinite-dimensional. Finally, most of our results go through
also in quantum field theory (although the measurement postulates are,
as is usual, geared to non-relativistic quantum mechanics.) Indeed,
the geometric treatment sheds new light on the second quantization
procedure. Because the space of quantum states is itself a symplectic
manifold equipped with a K\"ahler structure, it turns out that one can
use (a natural infinite dimensional generalization of) the machinery
of geometric quantization to carry out quantization again. The
resulting theory is precisely the second quantized one. Thus, second
quantization is indeed $({\rm quantization})^2$!  (For details, see
\cite{thesis}.)

We will conclude by listing a few of the important open problems.
First, we have given an intrinsically geometric formulation of the
five postulates of quantum mechanics that deal with kinematics,
unitary evolution and measurements of observables (possibly with
continuous spectra). However, we did not include the spin-statistics
postulate. The reason is that we do not have a succinct formulation of
this postulate which refers only to the essential geometric
structures. Obtaining such a formulation is a key open problem. The
remaining issues deal with generalizations of the standard
framework. We saw that it is rather straightforward to extend dynamics
by allowing the Hamiltonian to be any densely-defined function on the
quantum phase space. However, unless it is an observable function in
the sense of definition \ref{defn_observable_fn}, we may not have a
consistent measurement theory for it.  Whether this is a problem is
not so clear. Indeed, this feature arises already for the non-linear
Schr\"odinger equations considered in the literature and there it is
generally not perceived as a problem.  However, a systematic analysis
of this issue should be carried out. Next, as indicated in section
\ref{sec3.B}, there is possibility that there exist
infinite-dimensional K\"ahler manifolds with constant holomorphic
sectional curvature ($=2/\hbar$) which are {\it not} isomorphic to a
projective Hilbert space. If this does turn out to be the case, we
will obtain viable, non-trivial generalizations of quantum kinematics
for which even the measurement theory could be developed in
detail. Therefore, it is important to settle this issue. Also, even in
the finite-dimensional case, we do not know if there exist {\it any}
K\"ahler manifolds other than projective Hilbert spaces for which a
satisfactory measurement theory can be developed.  Even isolated
examples of such manifolds would be very illuminating. The final issue
stems from the fact that the space of quantum states shares several
features with the classical phase space. Is there then a quantization
procedure to arrive directly at $(\P, \w, g)$ without having to pass
through the Hilbert space? Not only may the answer be in the
affirmative but the new procedure may even be useful in cases when the
standard quantization procedure encounters difficulties. That is, such
a procedure may itself suggest generalizations of quantum mechanics.

\acknowledgements
\vskip20pt
We would like to thank Lane Hughston for correspondence, Ted Newman
for suggesting that we examine the WKB approximation and Domenico
Guilini for pointing out some early references.  This work was
supported in part by the NSF grants PHY93-96246 and PHY95-14240 and by
the Eberly fund of The Pennsylvania State University.


\end{document}